\documentclass[a4paper,12pt]{article}

\title{A cut-free sequent calculus for the bi-intuitionistic logic \texttt{2Int}}
\author{Sara Ayhan\\
Ruhr University Bochum\\
sara.ayhan@rub.de}

\date{September 2020}

\usepackage[english]{babel}
\usepackage{xcolor}
\usepackage{enumerate}
\usepackage{natbib}
\usepackage{amsmath}
\usepackage{amsfonts}
\usepackage{amssymb}
\usepackage{amsthm}
\usepackage{calrsfs}
\usepackage{graphicx}
\usepackage{bussproofs}
\usepackage{proof}
\usepackage{latexsym}
\usepackage[ND,SEQ]{prftree}
\usepackage{multicol, caption}
\usepackage{setspace}
\usepackage{stmaryrd}
\usepackage{xpatch}
\makeatletter
\newcommand{\MSonehalfspacing}{%
  \setstretch{1.44}
  \ifcase \@ptsize \relax 
    \setstretch {1.448}%
  \or 
    \setstretch {1.399}%
  \or 
    \setstretch {1.433}%
  \fi
}

\makeatother
\MSonehalfspacing
\expandafter\def\expandafter\quote\expandafter{\quote\singlespacing}
\makeatletter
\renewcommand\@biblabel[1]{}
\makeatother

\usepackage[paper=a4paper,left=25mm,right=35mm,top=25mm,bottom=20mm]{geometry} 

\newtheorem{theorem}{Theorem}[subsection]

\newtheorem{definition}[theorem]{Definition}

\newtheorem{lemma}[theorem]{Lemma}

\makeatletter
\xpatchcmd{\@thm}{\thm@headpunct{.}}{\thm@headpunct{}}{}
\makeatother

\begin{document}

\maketitle

\section{Introduction}
The purpose of this paper is to introduce a bi-intuitionistic sequent calculus and to give proofs of admissibility for its structural rules. 
Since I will ponder over the philosophical problems and implications of this calculus in a different paper, I only want to make some brief comments on these matters here. 
The calculus I will present, called \texttt{SC2Int}, is a sequent calculus for the bi-intuitionistic logic \texttt{2Int}, which Wansing presents in \citeyearpar{Wansing2016a}.
There he also gives a natural deduction system for this logic, \texttt{N2Int}, to which \texttt{SC2Int} is equivalent in terms of what is derivable.
I will spell out below what this amounts to exactly.
What is important is that these calculi represent a kind of bilateralist reasoning, since they do not only internalize processes of verification or provability but also the dual processes in terms of falsification or what is called \emph{dual provability}.
In \citep{Wansing2017} a normal form theorem for \texttt{N2Int} is stated, here, I want to prove a cut-elimination theorem for \texttt{SC2Int}, i.e. if successful, this would extend the results existing so far.

\section{The calculus \texttt{SC2Int}}
The language $\mathcal{L}_{2Int}$ of \texttt{2Int}, as given by Wansing, is defined in Backus-Naur form as follows:\\
$A ::= p \mid \bot \mid \top \mid (A \wedge A) \mid (A \vee A) \mid (A \rightarrow A) \mid (A \Yleft A)$.\\
As can be seen, we have a non-standard connective in this language, namely the operator of co-implication $\Yleft$, which acts as a dual to implication, just like conjunction and disjunction can be seen as dual connectives.
With that, we are in the realms of so-called bi-intuitionistic logic, which is a conservative extension of intuitionistic logic with co-implication.\footnote{Note that there is also a use of \textit{bi-intuitionistic logic} in the literature to refer to a specific system, namely \texttt{BiInt}, also called \textit{Heyting-Brouwer logic} (e.g. \citep{Rauszer, Gor, Postniece, Kowalski}). Co-implication is there to be understood to internalize the preservation of non-truth from the conclusion to the premises in a valid inference. The system 2Int, which is treated here, uses the same language as \texttt{BiInt}, but the meaning of co-implication differs (cf. \citep[p. 30f.]{Wansing2016a, Wansing2016b, Wansing2017}).}
We read $A \Yleft B$ as `B co-implies A'. 
\\
The general design of \texttt{SC2Int} resembles the intuitionistic sequent calculus \texttt{G3ip}.
The distinguishing features of this calculus consist in the shared contexts for all the logical rules, the axiom (in our calculus the reflexivity rules) being restricted to atomic formulas and the admissibility of \textit{all} structural rules (cf. \citep[p. 28-30]{Negri} for more information about the origins of this calculus).
Another distinguishing feature is the repetition of $A \rightarrow B$ in the left premise of the left introduction rule for implication, which is necessary for the proof of admissibility of contraction. 
Here, this happens in $\rightarrow L^{a}$ as well as with $A \Yleft B$ in $\Yleft L^{c}$.

We will use $p, q, r,...$ for atomic formulas, $A, B, C,...$ for arbitrary formulas, and $\Gamma, \Delta, \Gamma',...$ for multisets of formulas.
Sequents are of the form $(\Gamma; \Delta) \vdash^{*} C$ (with $\Gamma$ and $\Delta$ being finite, possibly empty multisets), which are read as ``From the verification of all formulas in $\Gamma$ and the falsification of all formulas in $\Delta$ one can derive the verification (resp. falsification) of $C$ for $*= +$ (resp. $* = -$)".  
Thus, we have a calculus in which a duality of derivability relations is considered, not only the one of verification but also the one of falsification.\footnote{In \texttt{N2Int} this is indicated by using single lines for verification and double lines for falsification.}
The formulas in $\Gamma$ can then be understood as \textit{assumptions}, while the formulas in $\Delta$ can be understood as \textit{counterassumptions}.
\texttt{SC2Int} is equivalent to \texttt{N2Int} in that we have a proof in \texttt{N2Int} of $A$ from the pair $(\Gamma; \Delta)$ of assumptions $\Gamma$ and counterassumptions $\Delta$, iff the sequent $(\Gamma; \Delta) \vdash^{+} A$ is derivable in \texttt{SC2Int} and we have a dual proof of $A$ from the pair $(\Gamma; \Delta)$ of assumptions $\Gamma$ and counterassumptions $\Delta$, iff the sequent $(\Gamma; \Delta) \vdash^{-} A$ is derivable in \texttt{SC2Int}.

In contrast to \texttt{G3ip}, there will be no distinction between axioms and logical rules but within the logical rules the zero-premise rules, which comprise $Rf^{+}$, $Rf^{-}$, $\bot L^{a}$, $\top L^{c}, \bot R^{-}$, and $\top R^{+}$, are distinguished from the non-zero-premise rules due to the special role of the former for the admissibility proofs below.
Each of the logical rules has a \textit{context} designated by $\Gamma$ and $\Delta$, \textit{active formulas} designated by $A$ and $B$ and a \textit{principal formula}, which is the one introduced on the left or right side of $\vdash^{*}$.
Within the right introduction rules we need to distinguish whether the derivability relation expresses verification or falsification by using the superscripts + and -.
Within the left rules this is not necessary, but what is needed here is distinguishing an introduction of the principal formula into the \textit{assumptions} from an introduction into the \textit{counterassumptions}.
The former are indexed by superscript $a$, while the latter are indexed by superscript $c$.
The set of $R^{+}$ and $L^{a}$ rules are the \emph{proof rules}; the set of $R^{-}$ and $L^{c}$ rules are the \emph{dual proof rules}.\\

\textbf{\large{SC2Int}}

\vspace{0.5cm}
\quad \quad \quad For $\ast \in$ \{+, -\}:

\vspace{0.2cm}

\quad  \quad \quad
\infer[\scriptstyle Rf^{+}]{(\Gamma, p; \Delta) \vdash^{+} p}{}
\quad \quad
\infer[\scriptstyle Rf^{-}]{(\Gamma; \Delta, p) \vdash^{-} p}{}

\vspace{0.4cm}
\quad  \quad \quad
\infer[\scriptstyle\bot L^{a}]{(\Gamma, \bot; \Delta) \vdash^{*} C}{}
\quad \quad
\infer[\scriptstyle\top L^{c}]{(\Gamma; \Delta, \top) \vdash^{*} C}{}

\vspace{0.4cm}
\quad  \quad \quad
\infer[\scriptstyle\bot R^{-}]{(\Gamma; \Delta) \vdash^{-} \bot}{}
\quad \quad \quad
\infer[\scriptstyle\top R^{+}]{(\Gamma; \Delta) \vdash^{+} \top}{}

\vspace{0.8cm}
\quad  \quad \quad
\infer[\scriptstyle\wedge R^{+}]{(\Gamma; \Delta) \vdash^{+} A \wedge B}{(\Gamma; \Delta) \vdash^{+} A \quad \quad (\Gamma; \Delta) \vdash^{+} B}
\quad  \quad
\infer[\scriptstyle\wedge L^{a}]{(\Gamma, A \wedge B; \Delta) \vdash^{*} C}{(\Gamma, A, B; \Delta) \vdash^{*} C}

\vspace{0.4cm}
\quad  \quad \quad
\infer[\scriptstyle\wedge R^{-}_{1}]{(\Gamma; \Delta) \vdash^{-} A \wedge B}{(\Gamma; \Delta) \vdash^{-} A}
\quad  \quad
\infer[\scriptstyle\wedge R^{-}_{2}]{(\Gamma; \Delta) \vdash^{-} A \wedge B}{(\Gamma; \Delta) \vdash^{-} B}

\vspace{0.4cm}
\quad  \quad \quad
\infer[\scriptstyle\wedge L^{c}]{(\Gamma; \Delta, A \wedge B) \vdash^{*} C}{(\Gamma; \Delta, A) \vdash^{*} C \quad \quad (\Gamma; \Delta, B) \vdash^{*} C}

\vspace{0.8cm}
\quad  \quad \quad
\infer[\scriptstyle\vee R^{+}_{1}]{(\Gamma; \Delta) \vdash^{+} A \vee B}{(\Gamma; \Delta) \vdash^{+} A}
\quad  \quad
\infer[\scriptstyle\vee R^{+}_{2}]{(\Gamma; \Delta) \vdash^{+} A \vee B}{(\Gamma; \Delta) \vdash^{+} B}

\vspace{0.4cm}
\quad  \quad \quad
\infer[\scriptstyle\vee L^{a}]{(\Gamma, A \vee B; \Delta) \vdash^{*} C}{(\Gamma, A; \Delta) \vdash^{*} C \quad \quad (\Gamma, B; \Delta) \vdash^{*} C}

\vspace{0.4cm}
\quad  \quad \quad
\infer[\scriptstyle\vee R^{-}]{(\Gamma; \Delta) \vdash^{-} A \vee B}{(\Gamma; \Delta) \vdash^{-} A \quad \quad (\Gamma; \Delta) \vdash^{-} B}
\quad  \quad
\infer[\scriptstyle\vee L^{c}]{(\Gamma; \Delta, A \vee B) \vdash^{*} C}{(\Gamma; \Delta, A, B) \vdash^{*} C}

\vspace{0.8cm}
\quad  \quad \quad
\infer[\scriptstyle\rightarrow R^{+}]{(\Gamma; \Delta) \vdash^{+} A \rightarrow B}{(\Gamma, A; \Delta) \vdash^{+} B}
\quad \quad
\infer[\scriptstyle\rightarrow L^{a}]{(\Gamma, A \rightarrow B; \Delta) \vdash^{*} C}{(\Gamma, A \rightarrow B; \Delta) \vdash^{+} A \quad \quad (\Gamma, B; \Delta) \vdash^{*} C}

\vspace{0.8cm}
\quad  \quad \quad
\infer[\scriptstyle\rightarrow R^{-}]{(\Gamma;\Delta) \vdash^{-} A \rightarrow B}{(\Gamma;\Delta)\vdash^{+} A \quad \quad (\Gamma;\Delta) \vdash^{-} B}
\quad \quad
\infer[\scriptstyle\rightarrow L^{c}]{(\Gamma; \Delta, A \rightarrow B) \vdash^{*} C}{(\Gamma, A; \Delta, B) \vdash^{*} C}

\vspace{0.8cm}
\quad  \quad \quad
\infer[\scriptstyle\Yleft R^{+}]{(\Gamma; \Delta) \vdash^{+} A \Yleft B}{(\Gamma; \Delta) \vdash^{+} A \quad \quad (\Gamma; \Delta) \vdash^{-} B}
\quad \quad
\infer[\scriptstyle\Yleft L^{a}]{(\Gamma, A \Yleft B;\Delta) \vdash^{*} C}{(\Gamma, A; \Delta, B) \vdash^{*} C}

\vspace{0.8cm}
\quad  \quad \quad
\infer[\scriptstyle\Yleft R^{-}]{(\Gamma; \Delta) \vdash^{-} A \Yleft B}{(\Gamma; \Delta, B) \vdash^{-} A}
\quad \quad
\infer[\scriptstyle\Yleft L^{c}]{(\Gamma; \Delta, A \Yleft B) \vdash^{*} C}{(\Gamma; \Delta, A \Yleft B) \vdash^{-} B \quad \quad (\Gamma; \Delta, A) \vdash^{*} C}

Note that the rules for $\wedge L^{a}$, $\vee L^{c}$, $\rightarrow L^{c}$ and $\Yleft L^{a}$ could also be given in the form of two rules, each with only one active formula $A$ or $B$, as it is for example done in Gentzen's original calculus for the left conjunction rule. 
We need this single rule formulation, however, in order to get the invertibility of these rules (cf. Lemma 3.3.1 below), which is important for the proof of admissibility of contraction.
As said above, the structural rules do not have to be taken as primitive in the calculus but can be shown to be admissible.
We want to consider rules for weakening, contraction and cut.
Due to the dual nature of the calculus, we need two rules for each of these rules:

\vspace{0.8cm}
\quad  
\infer[\scriptstyle W^{a}]{(\Gamma, A; \Delta) \vdash^{*} C}{(\Gamma;\Delta) \vdash^{*} C}
\quad \quad
\infer[\scriptstyle W^{c}]{(\Gamma;\Delta, A) \vdash^{*} C}{(\Gamma;\Delta) \vdash^{*} C}

\vspace{0.8cm}
\quad  
\infer[\scriptstyle C^{a}]{(\Gamma, A;\Delta) \vdash^{*} C}{(\Gamma, A, A; \Delta) \vdash^{*} C}
\quad \quad
\infer[\scriptstyle C^{c}]{(\Gamma; \Delta, A) \vdash^{*} C}{(\Gamma;\Delta, A, A) \vdash^{*} C}

\vspace{1cm}
\quad  
\infer[\scriptstyle Cut^{a}]{(\Gamma, \Gamma'; \Delta, \Delta') \vdash^{*} C}{(\Gamma;\Delta) \vdash^{+} D \quad \quad (\Gamma', D; \Delta') \vdash^{\ast} C}
\quad \quad
\infer[\scriptstyle Cut^{c}]{(\Gamma, \Gamma'; \Delta, \Delta') \vdash^{*} C}{(\Gamma;\Delta) \vdash^{-} D \quad \quad (\Gamma'; \Delta', D) \vdash^{*} C}

\section{Proving admissibility of the structural rules}

\subsection{Preliminaries}
The proofs of admissibility of the structural rules and especially of cut-elimination are conducted analogously to the respective proofs of \citet[p. 30-40]{Negri} for \texttt{G3ip}.
The proofs will use induction on \textbf{weight of formulas} and \textbf{height of derivations}.

\vspace{0.3cm}
\begin{definition} The weight w(A) of a formula A is defined inductively by\\
$w(\bot) = w(\top) = 0$,\\
$w(p) = 1$ for atoms $p$,\\
$w(A ~\# ~B) = w(A) + w(B) + 1$ for $\# \in \{\wedge, \vee, \rightarrow, \Yleft\}$. 
\end{definition}

\vspace{0.3cm}

\begin{definition} A derivation in \texttt{SC2Int} is either an instance of a zero-premise rule, or an application of a logical rule to derivations concluding its premises. The \emph{height of a derivation} is the greatest number of successive applications of rules in it, where zero-premise rules have height 0.
\end{definition}

\vspace{0.3cm}

First, I will show that the reflexivity rules can be generalized to instances with arbitrary formulas, not only atomic formulas.

\vspace{0.3cm}

\begin{lemma} The sequents $(\Gamma, C; \Delta) \vdash^{+} C$ \textit{and} $(\Gamma; \Delta, C) \vdash^{-} C$ are derivable for an arbitrary formula $C$ and arbitrary context $(\Gamma; \Delta)$.
\end{lemma}

\proof
The proof is by induction on weight of $C$. 
If $w(C) \leq 1$, we have the 19 cases listed below.
Note that for some of the derivations there is more than one possibility to derive the desired sequent and also some of the conclusions of zero-premise rules are conclusions of more than one of those rules.
I will just show one exemplary derivation for each case, since this is enough for the proof.

	$C = \bot$. Then $(\Gamma, C; \Delta) \vdash^{+} C$ is an instance of $\bot L^{a}$ and $(\Gamma; \Delta, C) \vdash^{-} C$ is an instance of $\bot R^{-}$.
	
	$C = \top$. Then $(\Gamma, C; \Delta) \vdash^{+} C$ is an instance of $\top R^{+}$ and $(\Gamma; \Delta, C) \vdash^{-} C$ is an instance of $\top L^{c}$.
	
	 $C = p$ for some atom $p$. Then $(\Gamma, C; \Delta) \vdash^{+} C$ is an instance of $Rf^{+}$ and $(\Gamma; \Delta, C) \vdash^{-} C$ is an instance of $Rf^{-}$.

	$C = \bot \wedge \bot$. Then $(\Gamma, C; \Delta) \vdash^{+} C$ and $(\Gamma; \Delta, C) \vdash^{-} C$ are derived by
	    \vspace{0.3cm}

\quad \quad \quad  
\infer[\scriptstyle\wedge R^{+}]{(\Gamma, \bot \wedge \bot; \Delta) \vdash^{+} \bot \wedge \bot} 
{\infer[\scriptstyle\bot L^{a}]{(\Gamma, \bot, \bot; \Delta) \vdash^{+} \bot \wedge \bot}{}}
\quad and \quad 
\infer[\scriptstyle\wedge R^{-}]{(\Gamma; \Delta, \bot \wedge \bot) \vdash^{-} \bot \wedge \bot} 
{\infer[\scriptstyle\bot R^{-}]{(\Gamma; \Delta, \bot \wedge \bot) \vdash^{-} \bot}{}}

$C = \bot \vee \bot$. Then $(\Gamma, C; \Delta) \vdash^{+} C$ and $(\Gamma; \Delta, C) \vdash^{-} C$ are derived by
	\vspace{0.3cm}

{\scriptsize
\quad  \hspace{-0.5cm}
\infer[\scriptstyle\vee L^{a}]{(\Gamma, \bot \vee \bot; \Delta) \vdash^{+} \bot \vee \bot}{{\infer[\scriptstyle \bot L^{a}]{(\Gamma, \bot; \Delta) \vdash^{+} \bot \vee \bot} {}}\quad \quad  {\infer[\scriptstyle \bot L^{a}]{(\Gamma, \bot; \Delta) \vdash^{+} \bot \vee \bot} {}}}
\quad and \quad
\infer[\scriptstyle\vee R^{-}]{(\Gamma; \Delta, \bot \vee \bot) \vdash^{-} \bot \vee \bot}{{\infer[\scriptstyle \bot R^{-}]{(\Gamma; \Delta, \bot \vee \bot) \vdash^{-} \bot} {}}\quad \quad  {\infer[\scriptstyle \bot R^{-}]{(\Gamma; \Delta, \bot \vee \bot) \vdash^{-} \bot} {}}}}

$C = \bot \rightarrow \bot$. Then $(\Gamma, C; \Delta) \vdash^{+} C$ and $(\Gamma; \Delta, C) \vdash^{-} C$ are derived by
	\vspace{0.3cm}
	
	\quad  \quad  \quad 
 \infer[\scriptstyle\rightarrow R^{+}]{(\Gamma, \bot \rightarrow \bot; \Delta) \vdash^{+} \bot \rightarrow \bot} 
{\infer[\scriptstyle\bot L^{a}]{(\Gamma, \bot \rightarrow \bot, \bot; \Delta) \vdash^{+} \bot}{}} 
\quad \quad  and \quad \quad 
\infer[\scriptstyle\rightarrow L^{c}]{(\Gamma; \Delta, \bot \rightarrow \bot) \vdash^{-} \bot \rightarrow \bot} 
{\infer[\scriptstyle\bot L^{a}]{(\Gamma, \bot; \Delta, \bot) \vdash^{-} \bot \rightarrow \bot}{}}
	
$C = \bot \Yleft \bot$. Then $(\Gamma, C; \Delta) \vdash^{+} C$ and $(\Gamma; \Delta, C) \vdash^{-} C$ are derived by
	\vspace{0.3cm}
	
	\quad  \quad  \quad 
\infer[\scriptstyle\Yleft L^{a}]{(\Gamma, \bot \Yleft \bot; \Delta) \vdash^{+} \bot \Yleft \bot} 
{\infer[\scriptstyle\bot L^{a}]{(\Gamma, \bot; \Delta, \bot) \vdash^{+} \bot \Yleft \bot}{}}
\quad \quad and \quad \quad 
\infer[\scriptstyle\Yleft R^{-}]{(\Gamma; \Delta, \bot \Yleft \bot) \vdash^{-} \bot \Yleft \bot} 
{\infer[\scriptstyle\bot R^{-}]{(\Gamma; \Delta, \bot \Yleft \bot, \bot) \vdash^{-} \bot}{}}	
	
$C = \bot \wedge \top$. Then $(\Gamma, C; \Delta) \vdash^{+} C$ and $(\Gamma; \Delta, C) \vdash^{-} C$ are derived by
	
	\vspace{0.3cm}

\quad  \quad  \quad 
\infer[\scriptstyle\wedge L^{a}]{(\Gamma, \bot \wedge \top; \Delta) \vdash^{+} \bot \wedge \top} 
{\infer[\scriptstyle\bot L^{a}]{(\Gamma, \bot, \top; \Delta) \vdash^{+} \bot \wedge \top}{}}
\quad \quad and \quad \quad 
\infer[\scriptstyle\wedge R^{-}_{1}]{(\Gamma; \Delta, \bot \wedge \top) \vdash^{-} \bot \wedge \top} 
{\infer[\scriptstyle\bot R^{-}]{(\Gamma; \Delta, \bot \wedge \top) \vdash^{-} \bot}{}}

$C = \bot \vee \top$. Then $(\Gamma, C; \Delta) \vdash^{+} C$ and $(\Gamma; \Delta, C) \vdash^{-} C$ are derived by
	
	\vspace{0.3cm}

\quad  \quad  \quad 
\infer[\scriptstyle\vee R^{+}_{2}]{(\Gamma, \bot \vee \top; \Delta) \vdash^{+} \bot \vee \top} 
{\infer[\scriptstyle\top R^{+}]{(\Gamma, \bot \vee \top; \Delta) \vdash^{+} \top}{}}
\quad \quad and \quad \quad 
\infer[\scriptstyle\vee L^{c}]{(\Gamma; \Delta, \bot \vee \top) \vdash^{-} \bot \vee \top} 
{\infer[\scriptstyle\top L^{c}]{(\Gamma; \Delta, \bot, \top) \vdash^{-} \bot \vee \top}{}}

$C = \bot \rightarrow \top$. Then $(\Gamma, C; \Delta) \vdash^{+} C$ and $(\Gamma; \Delta, C) \vdash^{-} C$ are derived by
	
	\vspace{0.3cm}

\quad  \quad  \quad 
\infer[\scriptstyle\rightarrow R^{+}]{(\Gamma, \bot \rightarrow \top; \Delta) \vdash^{+} \bot \rightarrow \top} 
{\infer[\scriptstyle\top R^{+}]{(\Gamma, \bot \rightarrow \top, \bot; \Delta) \vdash^{+} \top}{}}
\quad \quad and \quad \quad 
\infer[\scriptstyle\rightarrow L^{c}]{(\Gamma; \Delta, \bot \rightarrow \top) \vdash^{-} \bot \rightarrow \top} 
{\infer[\scriptstyle\top L^{c}]{(\Gamma, \bot; \Delta, \top) \vdash^{-} \bot \rightarrow \top}{}}

$C = \bot \Yleft \top$. Then $(\Gamma, C; \Delta) \vdash^{+} C$ and $(\Gamma; \Delta, C) \vdash^{-} C$ are derived by
	
	\vspace{0.3cm}

\quad  \quad  \quad 
\infer[\scriptstyle\Yleft L^{a}]{(\Gamma, \bot \Yleft \top; \Delta) \vdash^{+} \bot \Yleft \top} 
{\infer[\scriptstyle\bot L^{a}]{(\Gamma, \bot; \Delta, \top) \vdash^{+} \bot \Yleft \top}{}}
\quad \quad and \quad \quad 
\infer[\scriptstyle\Yleft R^{-}]{(\Gamma; \Delta, \bot \Yleft \top) \vdash^{-} \bot \Yleft \top} 
{\infer[\scriptstyle\top L^{c}]{(\Gamma; \Delta, \bot \Yleft \top, \top) \vdash^{-} \bot}{}}

$C = \top \wedge \bot$. Then $(\Gamma, C; \Delta) \vdash^{+} C$ and $(\Gamma; \Delta, C) \vdash^{-} C$ are derived by
	
	\vspace{0.3cm}

\quad  \quad  \quad 
\infer[\scriptstyle\wedge L^{a}]{(\Gamma, \top \wedge \bot; \Delta) \vdash^{+} \top \wedge \bot} 
{\infer[\scriptstyle\bot L^{a}]{(\Gamma, \top, \bot; \Delta) \vdash^{+} \top \wedge \bot}{}}
\quad \quad and \quad \quad 
\infer[\scriptstyle\wedge R^{-}_{2}]{(\Gamma; \Delta, \top \wedge \bot) \vdash^{-} \top \wedge \bot} 
{\infer[\scriptstyle\bot R^{-}]{(\Gamma; \Delta, \top \wedge \bot) \vdash^{-} \bot}{}}

$C = \top \vee \bot$. Then $(\Gamma, C; \Delta) \vdash^{+} C$ and $(\Gamma; \Delta, C) \vdash^{-} C$ are derived by
	
	\vspace{0.3cm}

\quad  \quad  \quad 
\infer[\scriptstyle\vee R^{+}_{1}]{(\Gamma, \top \vee \bot; \Delta) \vdash^{+} \top \vee \bot} 
{\infer[\scriptstyle\top R^{+}]{(\Gamma, \top \vee \bot; \Delta) \vdash^{+} \top}{}}
\quad \quad and \quad \quad 
\infer[\scriptstyle\vee L^{c}]{(\Gamma; \Delta, \top \vee \bot) \vdash^{-} \top \vee \bot} 
{\infer[\scriptstyle\top L^{c}]{(\Gamma; \Delta, \top, \bot) \vdash^{-} \top \vee \bot}{}}		

$C = \top \rightarrow \bot$. Then $(\Gamma, C; \Delta) \vdash^{+} C$ and $(\Gamma; \Delta, C) \vdash^{-} C$ are derived by
	\vspace{0.3cm}

{\scriptsize
\quad  \hspace{-0.5cm}
\infer[\scriptstyle\rightarrow L^{a}]{(\Gamma, \top \rightarrow \bot; \Delta) \vdash^{+} \top \rightarrow \bot}{{\infer[\scriptstyle \top R^{+}]{(\Gamma, \top \rightarrow \bot; \Delta) \vdash^{+} \top} {}}\quad \quad  {\infer[\scriptstyle \bot L^{a}]{(\Gamma, \bot; \Delta) \vdash^{+} \top \rightarrow \bot} {}}}
\quad and \quad
\infer[\scriptstyle\rightarrow R^{-}]{(\Gamma; \Delta, \top \rightarrow \bot) \vdash^{-} \top \rightarrow \bot}{{\infer[\scriptstyle \top R^{+}]{(\Gamma; \Delta, \top \rightarrow \bot) \vdash^{+} \top} {}}\quad \quad  {\infer[\scriptstyle \bot R^{-}]{(\Gamma; \Delta, \top \rightarrow \bot) \vdash^{-} \bot} {}}}}

$C = \top \Yleft \bot$. Then $(\Gamma, C; \Delta) \vdash^{+} C$ and $(\Gamma; \Delta, C) \vdash^{-} C$ are derived by
	\vspace{0.3cm}

{\scriptsize
\quad  \hspace{-0.5cm}
\infer[\scriptstyle\Yleft R^{+}]{(\Gamma, \top \Yleft \bot; \Delta) \vdash^{+} \top \Yleft \bot}{{\infer[\scriptstyle \top R^{+}]{(\Gamma, \top \Yleft \bot; \Delta) \vdash^{+} \top} {}}\quad \quad  {\infer[\scriptstyle \bot L^{a}]{(\Gamma, \top \Yleft \bot; \Delta) \vdash^{-} \bot} {}}}
\quad and \quad
\infer[\scriptstyle\Yleft L^{c}]{(\Gamma; \Delta, \top \Yleft \bot) \vdash^{-} \top \Yleft \bot}{{\infer[\scriptstyle \bot R^{-}]{(\Gamma; \Delta, \top \Yleft \bot) \vdash^{-} \bot} {}}\quad \quad  {\infer[\scriptstyle \top L^{c}]{(\Gamma; \Delta, \top) \vdash^{-} \top \Yleft \bot} {}}}}

$C = \top \wedge \top$. Then $(\Gamma, C; \Delta) \vdash^{+} C$ and $(\Gamma; \Delta, C) \vdash^{-} C$ are derived by
	\vspace{0.3cm}

{\scriptsize
\quad  \hspace{-0.5cm}
\infer[\scriptstyle\wedge R^{+}]{(\Gamma, \top \wedge \top; \Delta) \vdash^{+} \top \wedge \top}{{\infer[\scriptstyle \top R^{+}]{(\Gamma, \top \wedge \top; \Delta) \vdash^{+} \top} {}}\quad \quad  {\infer[\scriptstyle \top R^{+}]{(\Gamma, \top \wedge \top; \Delta) \vdash^{+} \top} {}}}
\quad and \quad
\infer[\scriptstyle\wedge L^{c}]{(\Gamma; \Delta, \top \wedge \top) \vdash^{-} \top \wedge \top}{{\infer[\scriptstyle \top L^{c}]{(\Gamma; \Delta, \top) \vdash^{-} \top \wedge \top} {}}\quad \quad  {\infer[\scriptstyle \top L^{c}]{(\Gamma; \Delta, \top) \vdash^{-} \top \wedge \top} {}}}}

$C = \top \vee \top$. Then $(\Gamma, C; \Delta) \vdash^{+} C$ and $(\Gamma; \Delta, C) \vdash^{-} C$ are derived by
	
	\vspace{0.3cm}

\quad  \quad  \quad 
\infer[\scriptstyle\vee R^{+}]{(\Gamma, \top \vee \top; \Delta) \vdash^{+} \top \vee \top} 
{\infer[\scriptstyle\top R^{+}]{(\Gamma, \top \vee \top; \Delta) \vdash^{+} \top}{}}
\quad \quad and \quad \quad 
\infer[\scriptstyle\vee L^{c}]{(\Gamma; \Delta, \top \vee \top) \vdash^{-} \top \vee \top} 
{\infer[\scriptstyle\top L^{c}]{(\Gamma; \Delta, \top, \top) \vdash^{-} \top \vee \top}{}}

$C = \top \rightarrow \top$. Then $(\Gamma, C; \Delta) \vdash^{+} C$ and $(\Gamma; \Delta, C) \vdash^{-} C$ are derived by
	
	\vspace{0.3cm}

\quad  \quad  \quad 
\infer[\scriptstyle\rightarrow R^{+}]{(\Gamma, \top \rightarrow \top; \Delta) \vdash^{+} \top \rightarrow \top} 
{\infer[\scriptstyle\top R^{+}]{(\Gamma, \top \rightarrow \top, \top; \Delta) \vdash^{+} \top}{}}
\quad \quad and \quad \quad 
\infer[\scriptstyle\rightarrow L^{c}]{(\Gamma; \Delta, \top \rightarrow \top) \vdash^{-} \top \rightarrow \top} 
{\infer[\scriptstyle\top L^{c}]{(\Gamma, \top; \Delta, \top) \vdash^{-} \top \rightarrow \top}{}}

$C = \top \Yleft \top$. Then $(\Gamma, C; \Delta) \vdash^{+} C$ and $(\Gamma; \Delta, C) \vdash^{-} C$ are derived by
	
	\vspace{0.3cm}

\quad  \quad  \quad 
\infer[\scriptstyle\Yleft L^{a}]{(\Gamma, \top \Yleft \top; \Delta) \vdash^{+} \top \Yleft \top} 
{\infer[\scriptstyle\top L^{c}]{(\Gamma, \top; \Delta, \top) \vdash^{+} \top \Yleft \top}{}}
\quad \quad and \quad \quad 
\infer[\scriptstyle\Yleft R^{-}]{(\Gamma; \Delta, \top \Yleft \top) \vdash^{-} \top \Yleft \top} 
{\infer[\scriptstyle\top L^{c}]{(\Gamma; \Delta, \top \Yleft \top, \top) \vdash^{-} \top}{}}

\vspace{0.8cm}
The inductive hypothesis is that $(\Gamma, C; \Delta) \vdash^{+} C$ and $(\Gamma; \Delta, C) \vdash^{-} C$ are derivable for all formulas $C$ with $w(C) \leq n$, and we have to show that $(\Gamma, D; \Delta) \vdash^{+} D$ and $(\Gamma; \Delta, D) \vdash^{-} D$ are derivable for formulas $D$ of weight $\leq n + 1$. 
There are four cases:

\vspace{0.3cm}
$D = A \wedge B$. By the definition of weight and our inductive hypothesis, $w(A) \leq n$ and $w(B) \leq n$. \\
We can derive $(\Gamma, A \wedge B; \Delta) \vdash^{+} A \wedge B$ by
\vspace{0.3cm}

\quad  \quad  \quad 
\infer[\scriptstyle\wedge R^{+}]{(\Gamma, A \wedge B; \Delta) \vdash^{+} A \wedge B}{\infer[\scriptstyle\wedge L^{a}]{(\Gamma, A \wedge B; \Delta) \vdash^{+} A} {(\Gamma, A, B; \Delta) \vdash^{+} A}\quad \quad \infer[\scriptstyle\wedge L^{a}]{(\Gamma, A \wedge B; \Delta) \vdash^{+} B} {(\Gamma, A, B; \Delta) \vdash^{+} B} }

and $(\Gamma; \Delta, A \wedge B) \vdash^{-} A \wedge B$ by

\vspace{0.3cm}

\quad  \quad  \quad 
\infer[\scriptstyle\wedge L^{c}]{(\Gamma; \Delta, A \wedge B) \vdash^{-} A \wedge B}{\infer[\scriptstyle\wedge R_{1}^{-}]{(\Gamma; \Delta, A) \vdash^{-} A \wedge B} {(\Gamma; \Delta, A) \vdash^{-} A}\quad \quad \infer[\scriptstyle\wedge R_{2}^{-}]{(\Gamma; \Delta, B) \vdash^{-} A \wedge B} {(\Gamma; \Delta, B) \vdash^{-} B} }

$(\Gamma; \Delta, A) \vdash^{-} A$ and $(\Gamma; \Delta, B) \vdash^{-} B$ are derivable by the inductive hypothesis and since the context is arbitrary, so are $(\Gamma', A; \Delta) \vdash^{+} A$ and $(\Gamma'', B; \Delta) \vdash^{+} B$, for $\Gamma' = \Gamma, B$ and $\Gamma'' = \Gamma, A$.

\vspace{0.3cm}
$D = A \vee B$. As before, $w(A) \leq n$ and $w(B) \leq n$. \\
We can derive $(\Gamma, A \vee B; \Delta) \vdash^{+} A \vee B$ by

\vspace{0.3cm}

\quad  \quad  \quad 
\infer[\scriptstyle\vee L^{a}]{(\Gamma, A \vee B; \Delta) \vdash^{+} A \vee B}{\infer[\scriptstyle\vee R_{1}^{+}]{(\Gamma, A; \Delta) \vdash^{+} A \vee B} {(\Gamma, A; \Delta) \vdash^{+} A}\quad \quad \infer[\scriptstyle\vee R_{2}^{+}]{(\Gamma, B; \Delta) \vdash^{+} A \vee B} {(\Gamma, B; \Delta) \vdash^{+} B} }

and $(\Gamma; \Delta, A \vee B) \vdash^{-} A \vee B$ by
\vspace{0.3cm}

\quad  \quad  \quad 
\infer[\scriptstyle\vee R^{-}]{(\Gamma; \Delta, A \vee B) \vdash^{-} A \vee B}{\infer[\scriptstyle\vee L^{c}]{(\Gamma; \Delta, A \vee B) \vdash^{-} A} {(\Gamma; \Delta, A, B) \vdash^{-} A}\quad \quad \infer[\scriptstyle\vee L^{c}]{(\Gamma; \Delta, A \vee B) \vdash^{-} B} {(\Gamma; \Delta, A, B) \vdash^{-} B} }

Again, by inductive hypothesis we get the derivability of $(\Gamma, A; \Delta) \vdash^{+} A$ and $(\Gamma, B; \Delta) \vdash^{+} B$ and since the context is arbitrary, $(\Gamma; \Delta', A) \vdash^{-} A$ and $(\Gamma; \Delta'', B) \vdash^{-} B$ are derivable, for $\Delta' = \Delta, B$ and $\Delta'' = \Delta, A$.

\vspace{0.3cm}
$D = A \rightarrow B$. As before, $w(A) \leq n$ and $w(B) \leq n$. \\
We can derive $(\Gamma, A \rightarrow B; \Delta) \vdash^{+} A \rightarrow B$ by
\vspace{0.3cm}

\quad  \quad  \quad 
\infer[\scriptstyle\rightarrow R^{+}]{(\Gamma, A \rightarrow B; \Delta) \vdash^{+} A \rightarrow B} 
{\infer[\scriptstyle\rightarrow L^{a}]{(\Gamma, A, A \rightarrow B; \Delta) \vdash^{+} B}
{{(\Gamma, A, A \rightarrow B; \Delta) \vdash^{+} A} 
\quad \quad {(\Gamma, A, B; \Delta) \vdash^{+} B}}}

and $(\Gamma; \Delta, A \rightarrow B) \vdash^{-} A \rightarrow B$ by

\vspace{0.3cm}

\quad  \quad  \quad 
\infer[\scriptstyle\rightarrow L^{c}]{(\Gamma; \Delta, A \rightarrow B) \vdash^{-} A \rightarrow B} 
{\infer[\scriptstyle\rightarrow R^{-}]{(\Gamma, A; \Delta, B) \vdash^{-} A \rightarrow B}
{{(\Gamma, A; \Delta, B) \vdash^{+} A} 
\quad \quad {(\Gamma, A; \Delta, B) \vdash^{-} B}}}

The case of $(\Gamma, A, B; \Delta) \vdash^{+} B$ was already mentioned in the case of conjunction and with the same reasoning $(\Gamma', A; \Delta) \vdash^{+} A$ for $\Gamma' = \Gamma, A \rightarrow B$, $(\Gamma, A; \Delta') \vdash^{+} A$ for $\Delta' = \Delta, B$ as well as $(\Gamma'; \Delta, B) \vdash^{-} B$ for $\Gamma' = \Gamma, A$ are derivable.

\vspace{0.3cm}
$D = A \Yleft B$. As before, $w(A) \leq n$ and $w(B) \leq n$. \\
We can derive $(\Gamma, A \Yleft B; \Delta) \vdash^{+} A \Yleft B$ by

\vspace{0.3cm}

\quad  \quad  \quad 
\infer[\scriptstyle\Yleft L^{a}]{(\Gamma, A \Yleft B; \Delta) \vdash^{+} A \Yleft B} 
{\infer[\scriptstyle\Yleft R^{+}]{(\Gamma, A; \Delta, B) \vdash^{+} A \Yleft B}
{{(\Gamma, A; \Delta, B) \vdash^{+} A} 
\quad \quad {(\Gamma, A; \Delta, B) \vdash^{-} B}}}

and $(\Gamma; \Delta, A \Yleft B) \vdash^{-} A \Yleft B$ by

\vspace{0.3cm}

\quad  \quad  \quad 
\infer[\scriptstyle\Yleft R^{-}]{(\Gamma; \Delta, A \Yleft B) \vdash^{-} A \Yleft B} 
{\infer[\scriptstyle\Yleft L^{c}]{(\Gamma; \Delta, B, A \Yleft B) \vdash^{-} A}
{{(\Gamma; \Delta, B, A \Yleft B) \vdash^{-} B} 
\quad \quad {(\Gamma; \Delta, A, B) \vdash^{-} A}}}

With the same reasoning as above $(\Gamma; \Delta', B) \vdash^{-} B$ is derivable for $\Delta'=\Delta, A \Yleft B$ and all other cases are already dealt with above.  
\qed

\subsection{Admissibility of weakening}
I will now start with the proof of admissibility of weakening by induction on height of derivation.
The general procedure when proving admissibility of a rule with this is to prove it for applications of this rule to conclusions of zero-premise rules and then generalize by induction on the number of applications of the rule to arbitrary derivations.
Thus, we can assume that there is only one instance - as the last step in the derivation - of the rule in question.

\begin{theorem}[Height-preserving weakening] If $(\Gamma; \Delta) \vdash^{*} C$ is derivable with a height of derivation at most n, then $(\Gamma, D; \Delta) \vdash^{*} C$ and $(\Gamma; \Delta, D) \vdash^{*} C$ are derivable with a height of derivation at most $n$ for arbitrary D.
\end{theorem}

\proof
If $n = 0$, then $(\Gamma; \Delta) \vdash^{*} C$ is a zero-premise rule, which means that one of the following six cases holds.
$C$ is an atom and 1) a formula in $\Gamma$ with $*=+$ or 2) a formula in $\Delta$ with $*=-$.
Otherwise it can be the case that 3) $C$ is $\top$ with $*=+$ or 4) $C$ is $\bot$ with $*=-$.
Lastly, it could be that 5) $\bot$ is a formula in $\Gamma$ or 6) $\top$ a formula in $\Delta$.
In either case, $(\Gamma, D; \Delta) \vdash^{*} C$ \textit{and} $(\Gamma; \Delta, D) \vdash^{*} C$ are conclusions of the respective zero-premise rules.
Our inductive hypothesis is now that height-preserving weakening is admissible up to derivations of height $\leq n$.
Let $(\Gamma; \Delta) \vdash^{*} C$ be derivable with a height of derivation at most $n + 1$.\\
If the last rule applied is $\wedge L^{a}$, then $\Gamma = \Gamma', A \wedge B$ and the last step is

\vspace{0.3cm}
\quad  \quad
\infer[\scriptstyle\wedge L^{a}]{(\Gamma', A \wedge B; \Delta) \vdash^{*} C}{(\Gamma', A, B; \Delta) \vdash^{*} C}

So $(\Gamma', A, B; \Delta) \vdash^{*} C$ is derivable in $\leq n$ steps.
By inductive hypothesis, also $(\Gamma', A, B, D; \Delta) \vdash^{*} C$  and $(\Gamma', A, B; \Delta, D) \vdash^{*} C$ are derivable in $\leq n$ steps.
Thus, the application of $\wedge L^{a}$ gives a derivation of $(\Gamma', A \wedge B, D; \Delta) \vdash^{*} C$  and $(\Gamma', A \wedge B; \Delta, D) \vdash^{*} C$ in $\leq n + 1$ steps.\\
If the last rule applied is $\wedge L^{c}$, then $\Delta = \Delta', A \wedge B$ and the last step is

\vspace{0.3cm}
\quad \quad
\infer[\scriptstyle\wedge L^{c}]{(\Gamma; \Delta', A \wedge B) \vdash^{*} C}{(\Gamma; \Delta', A) \vdash^{*} C \quad \quad (\Gamma; \Delta', B) \vdash^{*} C}

So $(\Gamma; \Delta', A) \vdash^{*} C$ and $(\Gamma; \Delta', B) \vdash^{*} C$ are derivable in $\leq n$ steps.
By inductive hypothesis, also $(\Gamma, D; \Delta', A) \vdash^{*} C$, $(\Gamma; \Delta', A, D) \vdash^{*} C$, $(\Gamma, D; \Delta', B) \vdash^{*} C$ and $(\Gamma; \Delta', B, D) \vdash^{*} C$ are derivable in $\leq n$ steps.
Thus, the application of $\wedge L^{c}$ to the first and the third premise and to the second and the fourth premise gives a derivation of $(\Gamma, D; \Delta', A \wedge B) \vdash^{*} C$  and $(\Gamma; \Delta', A \wedge B, D) \vdash^{*} C$, respectively, in $\leq n + 1$ steps.\\
If the last rule applied is $\wedge R^{+}$, then $C = A \wedge B$ and the last step is

\vspace{0.3cm}
\quad  \quad \quad
\infer[\scriptstyle\wedge R^{+}]{(\Gamma; \Delta) \vdash^{+} A \wedge B}{(\Gamma; \Delta) \vdash^{+} A \quad \quad (\Gamma; \Delta) \vdash^{+} B}

So $(\Gamma; \Delta) \vdash^{+} A$ and $(\Gamma; \Delta) \vdash^{+} B$ are derivable in $\leq n$ steps.
By inductive hypothesis, also $(\Gamma, D; \Delta) \vdash^{+} A$, $(\Gamma; \Delta, D) \vdash^{+} A$, $(\Gamma, D; \Delta) \vdash^{+} B$ and $(\Gamma; \Delta, D) \vdash^{+} B$ are derivable in $\leq n$ steps.
Thus, the application of $\wedge R^{+}$ to the first and the third premise and to the second and the fourth premise gives a derivation of $(\Gamma, D; \Delta) \vdash^{+} A \wedge B$  and $(\Gamma; \Delta, D) \vdash^{+} A \wedge B$, respectively, in $\leq n + 1$ steps.\\
If the last rule applied is $\wedge R^{-}_{1}$, then $C = A \wedge B$ and the last step is

\vspace{0.3cm}
\quad  \quad \quad
\infer[\scriptstyle\wedge R^{-}_{1}]{(\Gamma; \Delta) \vdash^{-} A \wedge B}{(\Gamma; \Delta) \vdash^{-} A}

So $(\Gamma; \Delta) \vdash^{-} A$ is derivable in $\leq n$ steps.
By inductive hypothesis, also $(\Gamma, D; \Delta) \vdash^{-} A$ and $(\Gamma; \Delta, D) \vdash^{-} A$ are derivable in $\leq n$ steps.
Thus, the application of $\wedge R^{-}_{1}$ gives a derivation of $(\Gamma, D; \Delta) \vdash^{-} A \wedge B$  and $(\Gamma; \Delta, D) \vdash^{-} A \wedge B$ in $\leq n + 1$ steps.

For the other logical rules the same can be shown with similar steps.
\qed

Now I want to show one other thing related to weakening because we will need this result later in our proof for the admissibility of the cut rules, namely that for the special case that the weakening formula is $\top$ for $W^{a}$ and respectively $\bot$ for $W^{c}$, the weakening rules are invertible, i.e.:

\vspace{0.8cm}
\quad  \quad \quad
\infer[\scriptstyle W_{inv}^{\top}]{(\Gamma; \Delta) \vdash^{*} C}{(\Gamma, \top; \Delta) \vdash^{*} C}
\quad \quad
\infer[\scriptstyle W_{inv}^{\bot}]{(\Gamma; \Delta) \vdash^{*} C}{(\Gamma; \Delta, \bot) \vdash^{*} C}

\begin{lemma}[Special case of inverted weakening] If $(\Gamma, \top; \Delta) \vdash^{*} C$ or $(\Gamma; \Delta, \bot) \vdash^{*} C$ are derivable with a height of derivation at most n, then so is $(\Gamma; \Delta) \vdash^{*} C$.
\end{lemma}

\proof
If $n = 0$, then exactly the same reasoning as for \textbf{Theorem 3.2.1} can be applied here.\\
Now we assume height-preserving invertibility for these two special cases of weakening up to height $n$, and let $(\Gamma, \top; \Delta) \vdash^{*} C$ and $(\Gamma; \Delta, \bot) \vdash^{*} C$ be derivable with a height of derivation $\leq n + 1$.
The proof works correspondingly to the proof of height-preserving weakening above, I will show it for the case of the $\rightarrow L^{c}$-rule this time, just to choose one that is not familiar in `usual' calculi, but it works similar for all logical connectives and their rules.\\
If the last rule applied is $\rightarrow L^{c}$, then we have $\Delta = \Delta', A \rightarrow B$ and the last step is 

\vspace{0.3cm}
\quad   
\infer[\scriptstyle\rightarrow L^{c}]{(\Gamma, \top; \Delta', A \rightarrow B) \vdash^{*} C}{(\Gamma, A, \top; \Delta', B) \vdash^{*} C}
\quad \quad or respectively
\quad \quad
\infer[\scriptstyle\rightarrow L^{c}]{(\Gamma; \Delta', A \rightarrow B, \bot) \vdash^{*} C}{(\Gamma, A; \Delta', B, \bot) \vdash^{*} C}

\vspace{0.3cm}
So, $(\Gamma, A, \top; \Delta', B) \vdash^{*} C$ and $(\Gamma, A; \Delta', B, \bot) \vdash^{*} C$ are derivable in $\leq n$ steps. Then by inductive hypothesis, $(\Gamma, A; \Delta', B) \vdash^{*} C$ is derivable in $\leq n$ steps. If we apply $\rightarrow L^{c}$ to this, this gives us $(\Gamma; \Delta', A \rightarrow B) \vdash^{*} C$ in $\leq n + 1$ steps.
\qed

\subsection{Admissibility of contraction}
Before we can prove the admissibility of the contraction rules, we need to prove the following lemma about the invertibility of premises and conclusions of the logical rules for the left introduction of formulas. 
Note that for $\rightarrow L^{a}$ and $\Yleft L^{c}$ the invertibility only holds for the right premises.\footnote{\citet[p. 33]{Negri} give a counterexample for the implication rule. The analogous counterexamples for \texttt{SC2Int} would be the derivability of the sequents $(\bot \rightarrow \bot; \emptyset) \vdash^{+} \bot \rightarrow \bot$ and $(\emptyset; \top \Yleft \top) \vdash^{-} \top \Yleft \top$.}

\begin{lemma}[Inversion]

\begin{itemize}
	\item[$(i_{1})$] If $(\Gamma, A \wedge B; \Delta) \vdash^{*} C$ is derivable with a height of derivation at most n, then $(\Gamma, A, B; \Delta) \vdash^{*} C$ is derivable with a height of derivation at most n.
	\item[$(i_{2})$] If $(\Gamma; \Delta, A \wedge B) \vdash^{*} C$ is derivable with a height of derivation at most n, then $(\Gamma; \Delta, A) \vdash^{*} C$ and $(\Gamma; \Delta, B) \vdash^{*} C$ are derivable with a height of derivation at most n.
	\item[$(ii_{1})$] If $(\Gamma, A \vee B; \Delta) \vdash^{*} C$ is derivable with a height of derivation at most n, then $(\Gamma, A; \Delta) \vdash^{*} C$ and $(\Gamma, B; \Delta) \vdash^{*} C$ are derivable with a height of derivation at most n.
	\item[$(ii_{2})$] If $(\Gamma; \Delta, A \vee B) \vdash^{*} C$ is derivable with a height of derivation at most n, then $(\Gamma; \Delta, A, B) \vdash^{*} C$ is derivable with a height of derivation at most n.
	\item[$(iii_{1})$] If $(\Gamma, A \rightarrow B; \Delta) \vdash^{*} C$ is derivable with a height of derivation at most n, then $(\Gamma, B; \Delta) \vdash^{*} C$ is derivable with a height of derivation at most n.
	\item[$(iii_{2})$]If $(\Gamma; \Delta, A \rightarrow B) \vdash^{*} C$ is derivable with a height of derivation at most n, then $(\Gamma, A; \Delta, B) \vdash^{*} C$ is derivable with a height of derivation at most n.
	\item[$(iv_{1})$] If $(\Gamma, A \Yleft B; \Delta) \vdash^{*} C$ is derivable with a height of derivation at most n, then $(\Gamma, A; \Delta, B) \vdash^{*} C$ is derivable with a height of derivation at most n.
	\item[$(iv_{2})$] If $(\Gamma; \Delta, A \Yleft B) \vdash^{*} C$ is derivable with a height of derivation at most n, then $(\Gamma; \Delta, A) \vdash^{*} C$ is derivable with a height of derivation at most n.
\end{itemize} 
\end{lemma} 

\proof
The proof is by induction on $n$.\\
1.) If $(\Gamma, A~ \# ~B; \Delta) \vdash^{*} C$ with $\# \in \{\wedge, \vee, \rightarrow, \Yleft\}$ is the conclusion of a zero-premise rule, then so are $(\Gamma, A, B; \Delta) \vdash^{*} C$, $(\Gamma, A; \Delta) \vdash^{*} C$, $(\Gamma, B; \Delta) \vdash^{*} C$, $(\Gamma; \Delta, B) \vdash^{*} C$ since $A~ \# ~B$ is neither atomic nor $\bot$ nor $\top$.\\
Now we assume height-preserving inversion up to height $n$, and let $(\Gamma, A ~\#~ B; \Delta) \vdash^{*} C$ be derivable with a height of derivation $\leq n + 1$.
\begin{itemize}
	\item[$(i_{1})$]
	Either $ A \wedge B $ is principal in the last rule or not.
	If $A \wedge B$ is the principal formula, the premise $(\Gamma, A, B; \Delta) \vdash^{*} C$  has a derivation of height $n$.
	If $A \wedge B$ is not principal in the last rule, then there must be one or two premises $(\Gamma', A \wedge B; \Delta') \vdash^{*} C'$, $(\Gamma'', A \wedge B; \Delta'') \vdash^{*} C''$ with a height of derivation $\leq n$.
	Then, by inductive hypothesis, also $(\Gamma', A, B; \Delta') \vdash^{*} C'$, $(\Gamma'', A, B; \Delta'') \vdash^{*} C''$ are derivable with a height of derivation $\leq n$.
	Now the last rule can be applied to these premises to conclude $(\Gamma, A, B; \Delta) \vdash^{*} C$ in at most $n+1$ steps.
	
	\item[$(ii_{1})$] Either $A\vee B$ is principal in the last rule or not.
	If $A \vee B$ is the principal formula, the premises $(\Gamma, A; \Delta) \vdash^{*} C$ and $(\Gamma, B; \Delta) \vdash^{*} C$ have a derivation of height $\leq n$.
	If $A \vee B$ is not principal in the last rule, then there must be one or two premises $(\Gamma', A \vee B; \Delta') \vdash^{*} C'$, $(\Gamma'', A \vee B; \Delta'') \vdash^{*} C''$ with a height of derivation $\leq n$.
	Then, by inductive hypothesis, also $(\Gamma', A; \Delta') \vdash^{*} C'$, $(\Gamma', B; \Delta') \vdash^{*} C'$ and $(\Gamma'', A; \Delta') \vdash^{*} C''$, $(\Gamma'', B; \Delta'') \vdash^{*} C''$ are derivable with a height of derivation $\leq n$.
	Now the last rule can be applied to the first and third premise to conclude $(\Gamma, A; \Delta) \vdash^{*} C$ and to the second and fourth premise to conclude $(\Gamma, B; \Delta) \vdash^{*} C$ in at most $n+1$ steps.
		\item[$(iii_{1})$]
	Either $A\rightarrow B$ is principal in the last rule or not.
	If $A \rightarrow B$ is the principal formula, the premise $(\Gamma, B; \Delta) \vdash^{*} C$  has a derivation of height $\leq n$.
	If $A \rightarrow B$ is not principal in the last rule, then there must be one or two premises $(\Gamma', A \rightarrow B; \Delta') \vdash^{*} C'$, $(\Gamma'', A \rightarrow B; \Delta'') \vdash^{*} C''$ with a height of derivation $\leq n$.
	Then, by inductive hypothesis, also $(\Gamma', B; \Delta') \vdash^{*} C'$, $(\Gamma'', B; \Delta'') \vdash^{*} C''$ are derivable with a height of derivation $\leq n$.
	Now the last rule can be applied to these premises to conclude $(\Gamma, B; \Delta) \vdash^{*} C$ in at most $n+1$ steps.
	\item[$(iv_{1})$]
	Either $A\Yleft B$ is principal in the last rule or not.
	If $A \Yleft B$ is the principal formula, then the premise $(\Gamma, A; \Delta, B) \vdash^{*} C$ has a derivation of height $n$.
	If $A \Yleft B$ is not principal in the last rule, then there must be one or two premises $(\Gamma', A \Yleft B; \Delta') \vdash^{*} C'$, $(\Gamma'', A \Yleft B; \Delta'') \vdash^{*} C''$ with a height of derivation $\leq n$.
	Then, by inductive hypothesis, also $(\Gamma', A; \Delta', B) \vdash^{*} C'$, $(\Gamma'', A; \Delta'', B) \vdash^{*} C''$ are derivable with a height of derivation $\leq n$.
	Now the last rule can be applied to these premises to conclude $(\Gamma, A; \Delta, B) \vdash^{*} C$ in at most $n+1$ steps.
\end{itemize}

2.) If $(\Gamma; \Delta, A~ \# ~B) \vdash^{*} C$ with $\# \in \{\wedge, \vee, \rightarrow, \Yleft\}$ is the conclusion of a zero-premise rule, then so are $(\Gamma; \Delta, A) \vdash^{*} C$, $(\Gamma; \Delta, B) \vdash^{*} C$, $(\Gamma; \Delta, A, B) \vdash^{*} C$, $(\Gamma, A; \Delta) \vdash^{*} C$ since $A~ \# ~B$ is neither atomic nor $\bot$ nor $\top$.\\
Now we assume height-preserving inversion up to height $n$, and let $(\Gamma; \Delta, A ~\#~ B) \vdash^{*} C$ be derivable with a height of derivation $\leq n+1$.
\begin{itemize}
	\item[$(i_{2})$]
	Either $A\wedge B$ is principal in the last rule or not.
	If $A \wedge B$ is the principal formula, the premises $(\Gamma; \Delta, A) \vdash^{*} C$ and $(\Gamma; \Delta, B) \vdash^{*} C$  have a derivation of height $\leq n$.
	If $A \wedge B$ is not principal in the last rule, then there must be one or two premises $(\Gamma'; \Delta', A \wedge B) \vdash^{*} C'$, $(\Gamma''; \Delta'', A \wedge B) \vdash^{*} C''$ with a height of derivation $\leq n$.
	Then, by inductive hypothesis, also $(\Gamma'; \Delta', A) \vdash^{*} C'$, $(\Gamma'; \Delta', B) \vdash^{*} C'$, $(\Gamma''; \Delta'', A) \vdash^{*} C''$, $(\Gamma''; \Delta'', B) \vdash^{*} C''$ are derivable with a height of derivation $\leq n$.
	Now the last rule can be applied to the first and third premise to conclude $(\Gamma; \Delta, A) \vdash^{*} C$ and to the second and fourth premise to conclude $(\Gamma; \Delta, B) \vdash^{*} C$ in at most $n+1$ steps.
	
	\item[$(ii_{2})$]
	Either $A\vee B$ is principal in the last rule or not.
	If $A \vee B$ is the principal formula, the premise $(\Gamma; \Delta, A, B) \vdash^{*} C$  has a derivation of height $n$.
	If $A \vee B$ is not principal in the last rule, then there must be one or two premises $(\Gamma'; \Delta', A \vee B) \vdash^{*} C'$, $(\Gamma''; \Delta'', A \vee B) \vdash^{*} C''$ with a height of derivation $\leq n$.
	Then, by inductive hypothesis, also $(\Gamma'; \Delta', A, B) \vdash^{*} C'$, $(\Gamma''; \Delta'', A, B) \vdash^{*} C''$ are derivable with a height of derivation $\leq n$.
	Now the last rule can be applied to these premises to conclude $(\Gamma; \Delta, A, B) \vdash^{*} C$ in at most $n+1$ steps.
	\item[$(iii_{2})$]
	Either $A\rightarrow B$ is principal in the last rule or not.
	If $A \rightarrow B$ is the principal formula, the premise $(\Gamma, A; \Delta, B) \vdash^{*} C$  has a derivation of height $n$.
	If $A \rightarrow B$ is not principal in the last rule, then there must be one or two premises $(\Gamma'; \Delta', A \rightarrow B) \vdash^{*} C'$, $(\Gamma''; \Delta'', A \rightarrow B) \vdash^{*} C''$ with a height of derivation $\leq n$.
	Then, by inductive hypothesis, also $(\Gamma', A; \Delta', B) \vdash^{*} C'$, $(\Gamma'', A; \Delta'', B) \vdash^{*} C''$ are derivable with a height of derivation $\leq n$.
	Now the last rule can be applied to these premises to conclude $(\Gamma, A; \Delta, B) \vdash^{*} C$ in at most $n+1$ steps.
	\item[$(iv_{2})$]
		Either $A\Yleft B$ is principal in the last rule or not.
	If $A \Yleft B$ is the principal formula, the premise $(\Gamma; \Delta, A) \vdash^{*} C$  has a derivation of height $\leq n$.
	If $A \Yleft B$ is not principal in the last rule, then there must be one or two premises $(\Gamma'; \Delta', A \Yleft B) \vdash^{*} C'$, $(\Gamma''; \Delta'', A \Yleft B) \vdash^{*} C''$ with a height of derivation $\leq n$.
	Then, by inductive hypothesis, also $(\Gamma'; \Delta', A) \vdash^{*} C'$, $(\Gamma''; \Delta'', A) \vdash^{*} C''$ are derivable with a height of derivation $\leq n$.
	Now the last rule can be applied to these premises to conclude $(\Gamma; \Delta, A) \vdash^{*} C$ in at most $n+1$ steps. 
	
\end{itemize}
\qed

Next, I will prove the admissibility of the contraction rules in \texttt{SC2Int}.

\begin{theorem}[Height-preserving contraction]
If $(\Gamma, D, D; \Delta) \vdash^{*} C$ is derivable with a height of derivation at most n, then $(\Gamma, D; \Delta) \vdash^{*} C$ is derivable with a height of derivation at most $n$ and if $(\Gamma; \Delta, D, D) \vdash^{*} C$ is derivable with a height of derivation at most n, then $(\Gamma; \Delta, D) \vdash^{*} C$ is derivable with a height of derivation at most $n$.
\end{theorem}

\proof
The proof is again by induction on the height of derivation $n$.\\
If $(\Gamma, D, D; \Delta) \vdash^{*} C$ (resp. $(\Gamma; \Delta, D, D) \vdash^{*} C$) is the conclusion of a zero-premise rule, then either C is an atom and contained in the antecedent, in the assumptions for $\vdash^{+}$ or in the counterassumptions for $\vdash^{-}$, or $\bot$ is part of the assumptions, or $\top$ is part of the counterassumptions, or $C=\top$ for $\vdash^{+}$, or $C=\bot$ for $\vdash^{-}$.
In either case, also $(\Gamma, D; \Delta) \vdash^{*} C$ (resp. $(\Gamma; \Delta, D) \vdash^{*} C$) is a conclusion of the respective zero-premise rule.\\
Let contraction be admissible up to derivation height $n$ and let $(\Gamma, D, D; \Delta) \vdash^{*} C$ (resp. $(\Gamma; \Delta, D, D) \vdash^{*} C$) be derivable in at most $n+1$ steps.
Either the contraction formula is not principal in the last inference step or it is principal.\\
If $D$ is not principal in the last rule concluding the premise of contraction $(\Gamma, D, D; \Delta)\\ \vdash^{*} C$, there must be one or two premises $(\Gamma', D, D; \Delta') \vdash^{*} C'$, $(\Gamma'', D, D; \Delta'') \vdash^{*} C'$ with a height of derivation $\leq n$.
So by inductive hypothesis, we can derive $(\Gamma', D; \Delta') \vdash^{*} C'$, $(\Gamma'', D; \Delta'') \vdash^{*} C'$ with a height of derivation $\leq n$.
Now the last rule can be applied to these premises to conclude $(\Gamma, D; \Delta) \vdash^{*} C$ in at most $n+1$ steps. 
For the case of $(\Gamma; \Delta, D, D) \vdash^{*} C$ being the premise of contraction, the same argument applies respectively.\\
If $D$ is principal in the last rule, we have to consider four cases for each contraction rule according to the form of $D$.
I will show the cases for $C^{c}$ this time; for $C^{a}$ the same arguments apply respectively.

\vspace{0.3cm}
$D=A\wedge B$. Then the last rule applied must be $\wedge L^{c}$ and we have as premises $(\Gamma; \Delta, A \wedge B, A) \vdash^{*} C$ and $(\Gamma; \Delta, A \wedge B, B) \vdash^{*} C$ with a derivation height $\leq n$.
By the inversion lemma this means that $(\Gamma; \Delta, A , A) \vdash^{*} C$ and $(\Gamma; \Delta, B, B) \vdash^{*} C$ are also derivable with a derivation height $\leq n$.
Then by inductive hypothesis, we get $(\Gamma; \Delta , A) \vdash^{*} C$ and $(\Gamma; \Delta, B) \vdash^{*} C$ with a height of derivation $\leq n$ and by applying $\wedge L^{c}$ we can derive $(\Gamma; \Delta, A \wedge B) \vdash^{*} C$ in at most $n+1$ steps.

\vspace{0.3cm}
$D=A\vee B$. Then the last rule applied must be $\vee L^{c}$ and $(\Gamma; \Delta, A \vee B, A, B) \vdash^{*} C$ is derivable with a height of derivation $\leq n$.
By the inversion lemma, also $(\Gamma; \Delta, A , B, A, B) \vdash^{*} C$ is derivable with a derivation height $\leq n$.
Then by inductive hypothesis (applied twice), we get $(\Gamma; \Delta , A, B) \vdash^{*} C$  with a height of derivation $\leq n$ and by applying $\vee L^{c}$ we can derive $(\Gamma; \Delta, A \vee B) \vdash^{*} C$ in at most $n+1$ steps.

\vspace{0.3cm}
$D=A\rightarrow B$. Then the last rule applied must be $\rightarrow L^{c}$ and accordingly $(\Gamma, A; \Delta, B,\\ A \rightarrow B) \vdash^{*} C$ is derivable with a height of derivation $\leq n$.
By the inversion lemma, then also $(\Gamma, A, A; \Delta, B, B) \vdash^{*} C$ is derivable with a derivation height $\leq n$.
By inductive hypothesis (applied twice), we get $(\Gamma, A; \Delta, B) \vdash^{*} C$ with a height of derivation $\leq n$ and by applying $\rightarrow L^{c}$ we can derive $(\Gamma; \Delta, A \rightarrow B) \vdash^{*} C$ in at most $n+1$ steps.

\vspace{0.3cm}
$D=A\Yleft B$. Then the last rule applied must be $\Yleft L^{c}$ and we have as premises $(\Gamma; \Delta, A \Yleft B, A \Yleft B) \vdash^{-} B$ and $(\Gamma; \Delta, A \Yleft B, A) \vdash^{*} C$ with a derivation height $\leq n$.
The inductive hypothesis applied to the first, gives us $(\Gamma; \Delta, A\Yleft B) \vdash^{-} B$ with a derivation height $\leq n$ and the inversion lemma applied to the second, also $(\Gamma; \Delta, A, A) \vdash^{*} C$ and again by inductive hypothesis $(\Gamma; \Delta, A) \vdash^{*} C$ with a derivation height $\leq n$.
By applying $\Yleft L^{c}$ we can now derive $(\Gamma; \Delta, A \Yleft B) \vdash^{*} C$ in at most $n+1$ steps. 
\qed

\subsection{Admissibility of cut}
Now, I will come to the main result, the proof of cut-elimination.
The proof shows that cuts can be permuted upward in a derivation until they reach one of the zero-premise rules the derivation started with.
When cut has reached zero-premise rules, the derivation can be transformed into one beginning with the conclusion of the cut, which can be shown by the following reasoning.

When both premises of cut are conclusions of a zero-premise rule, then the conclusion of cut is also a conclusion of one of these rules:
If the left premise is $(\Gamma, \bot; \Delta) \vdash^{*} D$, then the conclusion also has $\bot$ in the assumptions of the antecedent.
If the left premise is $(\Gamma; \Delta, \top) \vdash^{*} D$, then the conclusion also has $\top$ in the counterassumptions of the antecedent.
If the left premise of $Cut^{a}$ is $(\Gamma; \Delta) \vdash^{+} \top$ or the left premise of $Cut^{c}$ is $(\Gamma; \Delta) \vdash^{-} \bot$, then the right premise is $(\Gamma', \top; \Delta') \vdash^{*} C$ or $(\Gamma'; \Delta', \bot) \vdash^{*} C$ respectively.
These are conclusions of zero-premise rules only in one of the following cases:
\begin{itemize}
	\item $C$ is an atom in $\Gamma'$ for $\ast$ = + or $C$ is an atom in $\Delta'$ for $\ast$ = -
	\item $C = \top$ for $\ast$ = + or $C = \bot$ for $\ast$ = -
	\item $\bot$ is in $\Gamma'$ or $\top$ is in $\Delta'$
\end{itemize}
In each case the conclusion of cut $(\Gamma, \Gamma'; \Delta, \Delta') \vdash^{*} C$ is also a conclusion of the same zero-premise rule as the right premise. 
The last two possibilities are that the left premise is $(\Gamma, p; \Delta) \vdash^{+} p$ for $Cut^{a}$ or $(\Gamma; \Delta, p) \vdash^{-} p$ for $Cut^{c}$ respectively.
For the former case this means that the right premise is $(\Gamma', p; \Delta') \vdash^{*} C$. 
This is the conclusion of a zero-premise rule only in one of the following cases:
\begin{itemize}
	\item For $\ast$ = +: $C = p$, or $C$ is an atom in $\Gamma'$, or $C = \top$  
	\item For $\ast$ = -: $C$ is an atom in $\Delta'$, or $C = \bot$
	\item $\bot$ is in $\Gamma'$, or $\top$ is in $\Delta'$
\end{itemize}
In each case the conclusion of cut $(\Gamma, p, \Gamma'; \Delta, \Delta') \vdash^{*} C$ is also a conclusion of the same zero-premise rule as the right premise. 
For the latter case this means that the right premise is $(\Gamma'; \Delta', p) \vdash^{*} C$. 
This is the conclusion of a zero-premise rule only in one of the following cases:
\begin{itemize}
  \item For $\ast$ = +: $C$ is an atom in $\Gamma'$, or $C = \top$  
	\item For $\ast$ = -: $C = p$, or $C$ is an atom in $\Delta'$, or $C = \bot$
	\item $\bot$ is in $\Gamma'$, or $\top$ is in $\Delta'$
\end{itemize}
In each case the conclusion of cut $(\Gamma, \Gamma'; \Delta, p, \Delta') \vdash^{*} C$ is also a conclusion of the same zero-premise rule as the right premise. 
So, when cut has reached zero-premise rules as premises, the derivation can be transformed into one beginning with the conclusion of the cut by deleting the premises.

The proof is - as before - conducted in a manner corresponding to the proof of cut-elimination for \texttt{G3ip} by \citet{Negri}, which means that it is by induction on the weight of the cut formula and a subinduction on the cut-height, the sum of heights of derivations of the two premises of cut. 

\begin{definition}
The \emph{cut-height} of an application of one of the rules of cut in a derivation is the sum of heights of derivation of the two premises of the rule.
\end{definition}

In the proof permutations are given that always reduce the weight of the cut formula or the cut-height of instances of the rules.
When the cut formula is not principal in at least one (or both) of the premises of cut, cut-height is reduced.
In the other cases, i.e. in which the cut formula is principal in both premises, it is shown that cut-height and/or the weight of the cut formula can be reduced.
This process terminates since atoms cannot be principal formulas.

The difference between the height of a derivation and cut-height needs to be emphasized here, because it is essential to understand that if there are two instances of cut, one occurring below the other in the derivation, this does not necessarily mean that the lower instance has a greater cut-height than the upper. 
Let us suppose the upper instance of cut occurs in the derivation of the left premise of the lower cut.
The upper instance can have a cut-height which is greater than the height of either its premises because the \textit{sum} of the premises is what matters.
However, the lower instance can have as a right premise one with a much shorter derivation height than either of the premises of the upper cut, making the sum of the derivation heights of those two premises lesser than the one from the upper cut.
So, what follows is that it is not enough to show that occurrences of cut can be permuted upward in a derivation in order to show that cut-height decreases, but we need to calculate exactly the cut-height of each derivation in our proof.
As before, it can be assumed that in a given derivation the last instance is the one and only occurrence of cut.\\

\begin{theorem}
The cut rules 

\vspace{0.5cm}
\quad  \hspace{-0.5cm}
\infer[\scriptstyle Cut^{a}]{(\Gamma, \Gamma'; \Delta, \Delta') \vdash^{*} C}{(\Gamma;\Delta) \vdash^{+} D \quad \quad (\Gamma', D; \Delta') \vdash^{\ast} C}
\quad 
and
\quad 
\infer[\scriptstyle Cut^{c}]{(\Gamma, \Gamma'; \Delta, \Delta') \vdash^{*} C}{(\Gamma;\Delta) \vdash^{-} D \quad \quad (\Gamma'; \Delta', D) \vdash^{*} C} 

are admissible in \texttt{SC2Int}.
\end{theorem}

\proof
The proof is then organized as follows.
First, I consider the case that at least one premise in a cut is a conclusion of one of the zero-premise rules and show how cut can be eliminated in these cases.
Otherwise three cases can be distinguished: 1.) The cut formula is not principal in either premise of cut, 2.) the cut formula is principal in just one premise of cut, and 3.) the cut formula is principal in both premises of cut.

\section*{Cut with a conclusion of a zero-premise rule as premise}
\textbf{Cut with a conclusion of $Rf^{+}$, $Rf^{-}$, $\bot L^{a}$, $\top L^{c}, \bot R^{-}$, or $\top R^{+}$ as premise:}

If at least one of the premises of cut is a conclusion of one of the zero-premise rules, we distinguish three cases for both cut rules:

\subsection*{-1- Cut$^{a}$}
	\begin{enumerate} 
	
	\item[-1.1-] The left premise $(\Gamma;\Delta) \vdash^{+} D$ is a conclusion of a zero-premise-rule. There are four subcases:
	\begin{enumerate}
	
		\item The cut formula $D$ is an atom in $\Gamma$. Then the conclusion $(\Gamma, \Gamma'; \Delta, \Delta') \vdash^{*} C$ is derived from $(\Gamma', D; \Delta') \vdash^{\ast} C$ by $W^{a}$ and $W^{c}$.
		\item $\bot$ is a formula in $\Gamma$. Then $(\Gamma, \Gamma'; \Delta, \Delta') \vdash^{*} C$ is also a conclusion of $\bot L^{a}$.
		\item $\top$ is a formula in $\Delta$. Then $(\Gamma, \Gamma'; \Delta, \Delta') \vdash^{*} C$ is also a conclusion of $\top L^{c}$.
		\item $\top$ = D.  Then the right premise is $(\Gamma', \top; \Delta') \vdash^{*} C$ and $(\Gamma, \Gamma'; \Delta, \Delta') \vdash^{*} C$ follows by $W_{inv}^{\top}$ as well as $W^{a}$ and $W^{c}$.
	
	\end{enumerate}
	
	\item[-1.2-] The right premise $(\Gamma', D; \Delta') \vdash^{+} C$ is a conclusion of a zero-premise rule. There are six subcases:
	
	\begin{enumerate}
		\item $C$ is an atom in $\Gamma'$. Then $(\Gamma, \Gamma'; \Delta, \Delta') \vdash^{+} C$ is also a conclusion of $Rf^{+}$.
		\item $C$ = $D$. Then the left premise is $(\Gamma; \Delta) \vdash^{+} C$ and $(\Gamma, \Gamma'; \Delta, \Delta') \vdash^{+} C$ follows by $W^{a}$ and $W^{c}$.
		\item $\bot$ is in $\Gamma'$. Then $(\Gamma, \Gamma'; \Delta, \Delta') \vdash^{+} C$ is also a conclusion of $\bot L^{a}$.
		\item $\bot$ = D. Then the left premise is $(\Gamma; \Delta) \vdash^{+} \bot$ and is either a conclusion of $\bot L^{a}$ or $\top L^{c}$ (in which case cf. 1.1 (b) or 1.1 (c)) or it has been derived by a left rule. 
		There are eight cases according to the rule used which can be transformed into derivations with lesser cut-height.
		I will not show this here, since this is only a special case of the cases 3.1-3.8 below.
		\item $\top$ is in $\Delta'$. Then $(\Gamma, \Gamma'; \Delta, \Delta') \vdash^{+} C$ is also a conclusion of $\top L^{c}$.
		\item $\top$  = C. Then $(\Gamma, \Gamma'; \Delta, \Delta') \vdash^{+} C$ is also a conclusion of $\top R^{+}$.
	\end{enumerate}
	
	\item[-1.3-] The right premise $(\Gamma', D; \Delta') \vdash^{-} C$ is a conclusion of a zero-premise rule. There are five subcases:
	
	\begin{enumerate}
		\item $C$ is an atom in $\Delta'$. Then $(\Gamma, \Gamma'; \Delta, \Delta') \vdash^{-} C$ is also a conclusion of $Rf^{-}$.
		\item $\bot$ is in $\Gamma'$. Then $(\Gamma, \Gamma'; \Delta, \Delta') \vdash^{-} C$ is also a conclusion of $\bot L^{a}$.
		\item $\bot$ = D. Then the left premise is $(\Gamma; \Delta) \vdash^{+} \bot$ and the same as mentioned in 1.2 (d) holds.
		\item $\top$ is in $\Delta'$. Then $(\Gamma, \Gamma'; \Delta, \Delta') \vdash^{-} C$ is also a conclusion of $\top L^{c}$.
		\item $\bot$  = C. Then $(\Gamma, \Gamma'; \Delta, \Delta') \vdash^{-} C$ is also a conclusion of $\bot R^{-}$.
	\end{enumerate}
	\end{enumerate}

	\subsection*{-2- Cut$^{c}$}

\begin{enumerate} 
	\item[-2.1-] The left premise $(\Gamma;\Delta) \vdash^{-} D$ is a conclusion of a zero-premise rule. There are four subcases:
	\begin{enumerate}
	
		\item The cut formula $D$ is an atom in $\Delta$. Then the conclusion $(\Gamma, \Gamma'; \Delta, \Delta') \vdash^{*} C$ is derived from $(\Gamma'; \Delta', D) \vdash^{\ast} C$ by $W^{a}$ and $W^{c}$.
		\item $\bot$ is in $\Gamma$. Then $(\Gamma, \Gamma'; \Delta, \Delta') \vdash^{*} C$ is also a conclusion of $\bot L^{a}$.
		\item $\top$ is in $\Delta$. Then $(\Gamma, \Gamma'; \Delta, \Delta') \vdash^{*} C$ is also a conclusion of $\top L^{c}$.
		\item $\bot$ = D.  Then the right premise is $(\Gamma'; \Delta', \bot) \vdash^{*} C$ and $(\Gamma, \Gamma'; \Delta, \Delta') \vdash^{*} C$ follows by $W_{inv}^{\bot}$ as well as $W^{a}$ and $W^{c}$.
	
	\end{enumerate}
	
	\item[-2.2-] The right premise $(\Gamma'; \Delta', D) \vdash^{+} C$ is a conclusion of a zero-premise rule. There are five subcases:
	
	\begin{enumerate}
		\item $C$ is an atom in $\Gamma'$. Then $(\Gamma, \Gamma'; \Delta, \Delta') \vdash^{+} C$ is also a conclusion of $Rf^{+}$.
		\item $\bot$ is in $\Gamma'$. Then $(\Gamma, \Gamma'; \Delta, \Delta') \vdash^{+} C$ is also a conclusion of $\bot L^{a}$.
		\item $\top$ is in $\Delta'$. Then $(\Gamma, \Gamma'; \Delta, \Delta') \vdash^{+} C$ is also a conclusion of $\top L^{c}$.
		\item $\top$ = D. Then the left premise is $(\Gamma; \Delta) \vdash^{-} \top$ and the same as mentioned in 1.2 (d) holds.
		\item $\top$  = C. Then $(\Gamma, \Gamma'; \Delta, \Delta') \vdash^{+} C$ is also a conclusion of $\top R^{+}$.
	\end{enumerate}
	
	\item[-2.3-] The right premise $(\Gamma'; \Delta', D) \vdash^{-} C$ is a conclusion of a zero-premise rule. There are six subcases:
	
	\begin{enumerate}
		\item $C$ is an atom in $\Delta'$. Then $(\Gamma, \Gamma'; \Delta, \Delta') \vdash^{-} C$ is also a conclusion of $Rf^{-}$.
		\item $C$ = $D$. Then the left premise is $(\Gamma; \Delta) \vdash^{-} C$  and $(\Gamma, \Gamma'; \Delta, \Delta') \vdash^{-} C$ follows by $W^{a}$ and $W^{c}$.
		\item $\bot$ is in $\Gamma'$. Then $(\Gamma, \Gamma'; \Delta, \Delta') \vdash^{-} C$ is also a conclusion of $\bot L^{a}$.
		\item $\top$ is in $\Delta'$. Then $(\Gamma, \Gamma'; \Delta, \Delta') \vdash^{-} C$ is also a conclusion of $\top L^{c}$.
		\item $\top$ = D. Then the left premise is $(\Gamma; \Delta) \vdash^{-} \top$ and the same as mentioned in 1.2 (d) holds.
		\item $\bot$  = C. Then $(\Gamma, \Gamma'; \Delta, \Delta') \vdash^{-} C$ is also a conclusion of $\bot R^{-}$.
	\end{enumerate}
	\end{enumerate}

\section*{Cut with neither premise a conclusion of a zero-premise rule}
We distinguish the cases that a left rule is used to derive the left premise (cf. 3), a right rule is used to derive the left premise (cf. 5), a right or a left rule is used to derive the right premise with the cut formula not being principal there (cf. 4), and that a left rule is used to derive the right premise with the cut formula being principal (cf. 5). 
These cases can be subsumed in a more compact form as categorized below.
We assume, like \citet{Negri}, that in the derivations the topsequents, from left to right, have derivation heights $n$, $m$, $k$,...
\subsection*{-3- Cut not principal in the left premise}

If the cut formula $D$ is not principal in the left premise, this means that this premise is derived by a left introduction rule.
By permuting the order of the rules for the logical connectives with the cut rules, cut-height can be reduced in each of the following eight cases:
\begin{itemize}
	\item[-3.1-] $ \wedge L^{a} $ is the last rule used to derive the left premise with $\Gamma = \Gamma'', A \wedge B$. The derivations for $Cut^{a}$ and $Cut^{c}$ with cuts of cut-height $n+1+m$ are
	\vspace{0.3cm}

  \quad \hspace{-1.3cm}
{\footnotesize\infer[\scriptstyle Cut^{a}]{(\Gamma'', A \wedge B, \Gamma'; \Delta, \Delta') \vdash^{*} C}{\infer[\scriptstyle\wedge L^{a}]{(\Gamma'', A \wedge B; \Delta) \vdash^{+} D} {{(\Gamma'', A, B; \Delta) \vdash^{+} D}} \quad {(\Gamma', D; \Delta') \vdash^{*} C}\quad \quad {}}
\quad     
\infer[\scriptstyle Cut^{c}]{(\Gamma'', A \wedge B, \Gamma'; \Delta, \Delta') \vdash^{*} C}{\infer[\scriptstyle\wedge L^{a}]{(\Gamma'', A \wedge B; \Delta) \vdash^{-} D} {{(\Gamma'', A, B; \Delta) \vdash^{-} D}} \quad {(\Gamma'; \Delta', D) \vdash^{*} C}\quad \quad {}}}

These can be transformed into derivations with cuts of cut-height $n+m$:
\vspace{0.3cm}

\hspace{-1.3cm}
  \quad
{\footnotesize
\infer[\scriptstyle\wedge L^{a}]{(\Gamma'', A \wedge B, \Gamma'; \Delta, \Delta') \vdash^{*} C} 
{\infer[\scriptstyle Cut^{a}]{(\Gamma'', A, B, \Gamma'; \Delta, \Delta') \vdash^{*} C}
{{(\Gamma'', A, B; \Delta) \vdash^{+} D} 
\quad \quad {(\Gamma', D; \Delta') \vdash^{*} C}}}
\quad \quad
\infer[\scriptstyle\wedge L^{a}]{(\Gamma'', A \wedge B, \Gamma'; \Delta, \Delta') \vdash^{*} C} 
{\infer[\scriptstyle Cut^{c}]{(\Gamma'', A, B, \Gamma'; \Delta, \Delta') \vdash^{*} C}
{{(\Gamma'', A, B; \Delta) \vdash^{-} D} 
\quad \quad {(\Gamma'; \Delta', D) \vdash^{*} C}}}}

\item[-3.2-] $ \wedge L^{c} $ is the last rule used to derive the left premise with $\Delta = \Delta'', A \wedge B$. The derivations with cuts of cut-height $max (n,m)+1+k$ are
	\vspace{0.3cm}

\hspace{-1.3cm}
  \quad
{\tiny\infer[\scriptstyle Cut^{a}]{(\Gamma, \Gamma'; \Delta'', A \wedge B, \Delta') \vdash^{*} C}{\infer[\scriptstyle\wedge L^{c}]{(\Gamma; \Delta'', A \wedge B) \vdash^{+} D} {{(\Gamma; \Delta'', A) \vdash^{+} D} \quad {(\Gamma; \Delta'', B) \vdash^{+} D}} {(\Gamma', D; \Delta') \vdash^{*} C}\quad \quad \quad \quad \quad {}}   
\infer[\scriptstyle Cut^{c}]{(\Gamma, \Gamma'; \Delta'', A \wedge B, \Delta') \vdash^{*} C}{\infer[\scriptstyle\wedge L^{c}]{(\Gamma; \Delta'', A \wedge B) \vdash^{-} D} {{(\Gamma; \Delta'', A) \vdash^{-} D} \quad {(\Gamma; \Delta'', B) \vdash^{-} D}} {(\Gamma'; \Delta', D) \vdash^{*} C}\quad \quad \quad \quad \quad {}}}

These can be transformed into derivations each with two cuts of cut-height $n+k$ and $m+k$, respectively:
\vspace{0.3cm}

\hspace{-1.3cm}
  \quad
{\small\infer[\scriptstyle\wedge L^{c}]{(\Gamma, \Gamma'; \Delta'', A \wedge B, \Delta') \vdash^{*} C}{\infer[\scriptstyle Cut^{a}]{(\Gamma, \Gamma'; \Delta'', A, \Delta') \vdash^{*} C} {{(\Gamma; \Delta'', A) \vdash^{+} D} \quad {(\Gamma', D; \Delta') \vdash^{*} C}} \quad \quad {\infer[\scriptstyle Cut^{a}]{(\Gamma, \Gamma'; \Delta'', B, \Delta') \vdash^{*} C} {{(\Gamma; \Delta'', B) \vdash^{+} D} \quad {(\Gamma', D; \Delta') \vdash^{*} C}}}}   
\vspace{0.5cm}

\infer[\scriptstyle\wedge L^{c}]{(\Gamma, \Gamma'; \Delta'', A \wedge B, \Delta') \vdash^{*} C}{\infer[\scriptstyle Cut^{c}]{(\Gamma, \Gamma'; \Delta'', A, \Delta') \vdash^{*} C} {{(\Gamma; \Delta'', A) \vdash^{-} D} \quad {(\Gamma'; \Delta', D) \vdash^{*} C}} \quad \quad {\infer[\scriptstyle Cut^{c}]{(\Gamma, \Gamma'; \Delta'', B, \Delta') \vdash^{*} C} {{(\Gamma; \Delta'', B) \vdash^{-} D} \quad {(\Gamma'; \Delta', D) \vdash^{*} C}}}}}

\item[-3.3-] $ \vee L^{a} $ is the last rule used to derive the left premise with $\Gamma = \Gamma'', A \vee B$. The derivations with cuts of cut-height $max (n,m)+1+k$ are
	\vspace{0.3cm}

\hspace{-1.3cm}
  \quad \quad \quad \quad
{\small\infer[\scriptstyle Cut^{a}]{(\Gamma'', A \vee B, \Gamma'; \Delta, \Delta') \vdash^{*} C}{\infer[\scriptstyle\vee L^{a}]{(\Gamma'', A \vee B; \Delta) \vdash^{+} D} {{(\Gamma'', A; \Delta) \vdash^{+} D} \quad {(\Gamma'', B; \Delta) \vdash^{+} D}} {(\Gamma', D; \Delta') \vdash^{*} C}\quad \quad \quad \quad \quad {}}   
\vspace{0.1cm}

\quad
\infer[\scriptstyle Cut^{c}]{(\Gamma'', A \vee B, \Gamma'; \Delta, \Delta') \vdash^{*} C}{\infer[\scriptstyle\vee L^{a}]{(\Gamma'', A \vee B; \Delta) \vdash^{-} D} {{(\Gamma'', A; \Delta) \vdash^{-} D} \quad {(\Gamma'', B; \Delta) \vdash^{-} D}} {(\Gamma'; \Delta', D) \vdash^{*} C}\quad \quad \quad \quad \quad {}}}

These can be transformed into derivations each with two cuts of cut-height $n+k$ and $m+k$, respectively:
\vspace{0.3cm}

\hspace{-1.3cm}
  \quad \quad 
{\small\infer[\scriptstyle\vee L^{a}]{(\Gamma'', A \vee B, \Gamma'; \Delta, \Delta') \vdash^{*} C}{\infer[\scriptstyle Cut^{a}]{(\Gamma'', A, \Gamma'; \Delta, \Delta') \vdash^{*} C} {{(\Gamma'', A; \Delta) \vdash^{+} D} \quad {(\Gamma', D; \Delta') \vdash^{*} C}} \quad \quad {\infer[\scriptstyle Cut^{a}]{(\Gamma'', B, \Gamma'; \Delta, \Delta') \vdash^{*} C} {{(\Gamma'', B; \Delta) \vdash^{+} D} \quad {(\Gamma', D; \Delta') \vdash^{*} C}}}}   
\vspace{0.2cm}

\infer[\scriptstyle\vee L^{a}]{(\Gamma'', A \vee B, \Gamma'; \Delta, \Delta') \vdash^{*} C}{\infer[\scriptstyle Cut^{c}]{(\Gamma'', A, \Gamma'; \Delta, \Delta') \vdash^{*} C} {{(\Gamma'', A; \Delta) \vdash^{-} D} \quad {(\Gamma'; \Delta', D) \vdash^{*} C}} \quad \quad {\infer[\scriptstyle Cut^{c}]{(\Gamma'', B, \Gamma'; \Delta, \Delta') \vdash^{*} C} {{(\Gamma'', B; \Delta) \vdash^{-} D} \quad {(\Gamma'; \Delta', D) \vdash^{*} C}}}}}

\item[-3.4-] $ \vee L^{c} $ is the last rule used to derive the left premise with $\Delta = \Delta'', A \vee B$. The derivations with cuts of cut-height $n+1+m$ are
	\vspace{0.3cm}

\hspace{-1.3cm}
  \quad
{\footnotesize\infer[\scriptstyle Cut^{a}]{(\Gamma, \Gamma'; \Delta'', A \vee B, \Delta') \vdash^{*} C}{\infer[\scriptstyle\vee L^{c}]{(\Gamma; \Delta'', A \vee B) \vdash^{+} D} {{(\Gamma; \Delta'', A, B) \vdash^{+} D}} \quad {(\Gamma', D; \Delta') \vdash^{*} C}\quad \quad {}}
\quad     
\infer[\scriptstyle Cut^{c}]{(\Gamma, \Gamma'; \Delta'', A \vee B, \Delta') \vdash^{*} C}{\infer[\scriptstyle\vee L^{c}]{(\Gamma; \Delta'', A \vee B) \vdash^{-} D} {{(\Gamma; \Delta'', A, B) \vdash^{-} D}} \quad {(\Gamma'; \Delta', D) \vdash^{*} C}\quad \quad {}}}

These can be transformed into derivations with cuts of cut-height $n+m$:
\vspace{0.3cm}

\hspace{-1.3cm}
  \quad
{\footnotesize
\infer[\scriptstyle\vee L^{c}]{(\Gamma, \Gamma'; \Delta'', A \vee B, \Delta') \vdash^{*} C} 
{\infer[\scriptstyle Cut^{a}]{(\Gamma, \Gamma'; \Delta'', A, B, \Delta') \vdash^{*} C}
{{(\Gamma; \Delta'', A, B) \vdash^{+} D} 
\quad \quad {(\Gamma', D; \Delta') \vdash^{*} C}}}
\quad \quad
\infer[\scriptstyle\vee L^{c}]{(\Gamma, \Gamma'; \Delta'', A \vee B, \Delta') \vdash^{*} C} 
{\infer[\scriptstyle Cut^{c}]{(\Gamma, \Gamma'; \Delta'', A, B, \Delta') \vdash^{*} C}
{{(\Gamma; \Delta'', A, B) \vdash^{-} D} 
\quad \quad {(\Gamma'; \Delta', D) \vdash^{*} C}}}}

\item[-3.5-] $ \rightarrow L^{a} $ is the last rule used to derive the left premise with $\Gamma = \Gamma'', A \rightarrow B$. The derivations with cuts of cut-height $max (n,m)+1+k$ are
	\vspace{0.3cm}

\hspace{-1.3cm}
  \quad \quad \quad \quad
{\small\infer[\scriptstyle Cut^{a}]{(\Gamma'', A \rightarrow B, \Gamma'; \Delta, \Delta') \vdash^{*} C}{\infer[\scriptstyle\rightarrow L^{a}]{(\Gamma'', A \rightarrow B; \Delta) \vdash^{+} D} {{(\Gamma'', A \rightarrow B; \Delta) \vdash^{+} A} \quad {(\Gamma'', B; \Delta) \vdash^{+} D}} {(\Gamma', D; \Delta') \vdash^{*} C}\quad \quad \quad \quad \quad {}}  
\vspace{0.5cm}
 
\quad
\infer[\scriptstyle Cut^{c}]{(\Gamma'', A \rightarrow B, \Gamma'; \Delta, \Delta') \vdash^{*} C}{\infer[\scriptstyle\rightarrow L^{a}]{(\Gamma'', A \rightarrow B; \Delta) \vdash^{-} D} {{(\Gamma'', A \rightarrow B; \Delta) \vdash^{+} A} \quad {(\Gamma'', B; \Delta) \vdash^{-} D}}{(\Gamma'; \Delta', D) \vdash^{*} C}\quad \quad \quad \quad \quad {}}}

These can be transformed into derivations with cuts of cut-height $m+k$:
\vspace{0.3cm}

\hspace{-1.3cm}  \quad 
{\small\infer[\scriptstyle\rightarrow L^{a}]{(\Gamma'', A \rightarrow B, \Gamma'; \Delta, \Delta') \vdash^{*} C} {\infer[\scriptstyle W^{a/c}]{\quad \quad \quad {(\Gamma'', A \rightarrow B, \Gamma'; \Delta, \Delta') \vdash^{+} A}} {(\Gamma'', A \rightarrow B; \Delta) \vdash^{+} A} \quad \quad {\infer[\scriptstyle Cut^{a}]{(\Gamma'', B, \Gamma'; \Delta, \Delta') \vdash^{*} C} {{(\Gamma'', B; \Delta) \vdash^{+} D} \quad {(\Gamma', D; \Delta') \vdash^{*} C}}}}   
\vspace{0.1cm}

\infer[\scriptstyle\rightarrow L^{a}]{(\Gamma'', A \rightarrow B, \Gamma'; \Delta, \Delta') \vdash^{*} C}{\infer[\scriptstyle W^{a/c}]{{\quad \quad \quad(\Gamma'', A \rightarrow B, \Gamma'; \Delta, \Delta') \vdash^{+} A}} {{(\Gamma'', A \rightarrow B; \Delta) \vdash^{+} A} \quad } \quad \quad {\infer[\scriptstyle Cut^{c}]{(\Gamma'', B, \Gamma'; \Delta, \Delta') \vdash^{*} C} {{(\Gamma'', B; \Delta) \vdash^{-} D} \quad {(\Gamma'; \Delta', D) \vdash^{*} C}}}}}

\item[-3.6-] $ \rightarrow L^{c}$ is the last rule used to derive the left premise with $\Delta = \Delta'', A \rightarrow B$. The derivations with cuts of cut-height $n+1+m$ are
	\vspace{0.3cm}

\hspace{-1.3cm}
  \quad
{\footnotesize\infer[\scriptstyle Cut^{a}]{(\Gamma, \Gamma'; \Delta'', A \rightarrow B, \Delta') \vdash^{*} C}{\infer[\scriptstyle\rightarrow L^{c}]{(\Gamma; \Delta'', A \rightarrow B) \vdash^{+} D} {{(\Gamma, A; \Delta'', B) \vdash^{+} D}} \quad {(\Gamma', D; \Delta') \vdash^{*} C}\quad \quad {}}
\quad     
\infer[\scriptstyle Cut^{c}]{(\Gamma, \Gamma'; \Delta'', A \rightarrow B, \Delta') \vdash^{*} C}{\infer[\scriptstyle\rightarrow L^{c}]{(\Gamma; \Delta'', A \rightarrow B) \vdash^{-} D} {{(\Gamma, A; \Delta'', B) \vdash^{-} D}} \quad {(\Gamma'; \Delta', D) \vdash^{*} C}\quad \quad {}}}

These can be transformed into derivations with cuts of cut-height $n+m$:
\vspace{0.3cm}

\hspace{-1.3cm}
  \quad
{\footnotesize
\infer[\scriptstyle\rightarrow L^{c}]{(\Gamma, \Gamma'; \Delta'', A \rightarrow B, \Delta') \vdash^{*} C} 
{\infer[\scriptstyle Cut^{a}]{(\Gamma, A, \Gamma'; \Delta'', B, \Delta') \vdash^{*} C}
{{(\Gamma, A; \Delta'', B) \vdash^{+} D} 
\quad \quad {(\Gamma', D; \Delta') \vdash^{*} C}}}
\quad \quad
\infer[\scriptstyle\rightarrow L^{c}]{(\Gamma, \Gamma'; \Delta'', A \rightarrow B, \Delta') \vdash^{*} C} 
{\infer[\scriptstyle Cut^{c}]{(\Gamma, A, \Gamma'; \Delta'', B, \Delta') \vdash^{*} C}
{{(\Gamma, A; \Delta'', B) \vdash^{-} D} 
\quad \quad {(\Gamma'; \Delta', D) \vdash^{*} C}}}}

\item[-3.7-] $ \Yleft L^{a}$ is the last rule used to derive the left premise with $\Gamma = \Gamma'', A \Yleft B$. The derivations with cuts of cut-height $n+1+m$ are
	\vspace{0.3cm}

\hspace{-1.3cm}
  \quad
{\footnotesize\infer[\scriptstyle Cut^{a}]{(\Gamma'', A \Yleft B, \Gamma'; \Delta, \Delta') \vdash^{*} C}{\infer[\scriptstyle\Yleft L^{a}]{(\Gamma'', A \Yleft B; \Delta) \vdash^{+} D} {{(\Gamma'', A; \Delta, B) \vdash^{+} D}} \quad {(\Gamma', D; \Delta') \vdash^{*} C}\quad \quad {}}
\quad     
\infer[\scriptstyle Cut^{c}]{(\Gamma'', A \Yleft B, \Gamma'; \Delta, \Delta') \vdash^{*} C}{\infer[\scriptstyle\Yleft L^{a}]{(\Gamma'', A \Yleft B; \Delta) \vdash^{-} D} {{(\Gamma'', A; \Delta, B) \vdash^{-} D}} \quad {(\Gamma'; \Delta', D) \vdash^{*} C}\quad \quad {}}}

These can be transformed into derivations with cuts of cut-height $n+m$:
\vspace{0.3cm}

\hspace{-1.3cm}
  \quad
{\footnotesize
\infer[\scriptstyle\Yleft L^{a}]{(\Gamma'', A \Yleft B, \Gamma'; \Delta, \Delta') \vdash^{*} C} 
{\infer[\scriptstyle Cut^{a}]{(\Gamma'', A, \Gamma'; \Delta, B, \Delta') \vdash^{*} C}
{{(\Gamma'', A; \Delta, B) \vdash^{+} D} 
\quad \quad {(\Gamma', D; \Delta') \vdash^{*} C}}}
\quad \quad
\infer[\scriptstyle\Yleft L^{a}]{(\Gamma'', A \Yleft B, \Gamma'; \Delta, \Delta') \vdash^{*} C} 
{\infer[\scriptstyle Cut^{c}]{(\Gamma'', A, \Gamma'; \Delta, B, \Delta') \vdash^{*} C}
{{(\Gamma'', A; \Delta, B) \vdash^{-} D} 
\quad \quad {(\Gamma'; \Delta', D) \vdash^{*} C}}}}

\item[-3.8-] $ \Yleft L^{c} $ is the last rule used to derive the left premise with $\Delta = \Delta'', A \Yleft B$. The derivations with cuts of cut-height $max (n,m)+1+k$ are
	\vspace{0.3cm}

\hspace{-1.3cm}
  \quad \quad \quad \quad
{\small\infer[\scriptstyle Cut^{a}]{(\Gamma, \Gamma'; \Delta'', A \Yleft B, \Delta') \vdash^{*} C}{\infer[\scriptstyle\Yleft L^{c}]{(\Gamma; \Delta'', A \Yleft B) \vdash^{+} D} {{(\Gamma; \Delta'', A \Yleft B) \vdash^{-} B} \quad {(\Gamma; \Delta'', A) \vdash^{+} D}} {(\Gamma', D; \Delta') \vdash^{*} C}\quad \quad \quad \quad \quad {}}  
\vspace{0.5cm}
 
\quad
\infer[\scriptstyle Cut^{c}]{(\Gamma, \Gamma'; \Delta'', A \Yleft B, \Delta') \vdash^{*} C}{\infer[\scriptstyle\Yleft L^{c}]{(\Gamma; \Delta'', A \Yleft B) \vdash^{-} D} {{(\Gamma; \Delta'', A \Yleft B) \vdash^{-} B} \quad {(\Gamma; \Delta'', A) \vdash^{-} D}}{(\Gamma'; \Delta', D) \vdash^{*} C}\quad \quad \quad \quad \quad {}}}

These can be transformed into derivations with cuts of cut-height $m+k$:
\vspace{0.3cm}

\hspace{-1.3cm}
  \quad 
{\small\infer[\scriptstyle\Yleft L^{c}]{(\Gamma, \Gamma'; \Delta'', A \Yleft B, \Delta') \vdash^{*} C} {\infer[\scriptstyle W^{a/c}]{\quad \quad \quad {(\Gamma, \Gamma'; \Delta'', A \Yleft B, \Delta') \vdash^{-} B}} {(\Gamma; \Delta'', A \Yleft B) \vdash^{-} B} \quad \quad {\infer[\scriptstyle Cut^{a}]{(\Gamma, \Gamma'; \Delta'', A, \Delta') \vdash^{*} C} {{(\Gamma; \Delta'', A) \vdash^{+} D} \quad {(\Gamma', D; \Delta') \vdash^{*} C}}}}   
\vspace{0.1cm}

\infer[\scriptstyle\Yleft L^{c}]{(\Gamma, \Gamma'; \Delta'', A \Yleft B, \Delta') \vdash^{*} C}{\infer[\scriptstyle W^{a/c}]{{\quad \quad \quad(\Gamma, \Gamma'; \Delta'', A \Yleft B, \Delta') \vdash^{-} B}} {{(\Gamma; \Delta'', A \Yleft B) \vdash^{-} B} \quad } \quad \quad {\infer[\scriptstyle Cut^{c}]{(\Gamma, \Gamma'; \Delta'', A, \Delta') \vdash^{*} C} {{(\Gamma; \Delta'' A) \vdash^{-} D} \quad {(\Gamma'; \Delta', D) \vdash^{*} C}}}}}

\end{itemize}

As said above, cut-height is reduced in all cases.

\subsection*{-4- Cut formula \textit{D} principal in the left premise only}

The cases distinguished here concern the way the right premise is derived.
We can distinguish 16 cases and show for each case that the derivation of the right premise can be transformed into one containing only occurrences of cut with a reduced cut-height.

\begin{itemize}
	\item[-4.1-] $ \wedge L^{a} $ is the last rule used to derive the right premise with $\Gamma' = \Gamma'', A \wedge B$. The derivations with cuts of cut-height $n+m+1$ are
	\vspace{0.3cm}

\hspace{-1.3cm}
  \quad
{\footnotesize\infer[\scriptstyle Cut^{a}]{(\Gamma, \Gamma'', A \wedge B; \Delta, \Delta') \vdash^{*} C}{{(\Gamma; \Delta) \vdash^{+} D}\quad \quad {}\infer[\scriptstyle\wedge L^{a}]{(\Gamma'', A \wedge B, D; \Delta') \vdash^{*} C} {{(\Gamma'', A, B, D; \Delta') \vdash^{*} C}}}
\quad     
\infer[\scriptstyle Cut^{c}]{(\Gamma, \Gamma'', A \wedge B; \Delta, \Delta') \vdash^{*} C}{{(\Gamma; \Delta) \vdash^{-} D}\quad \quad {}\infer[\scriptstyle\wedge L^{a}]{(\Gamma'', A \wedge B; \Delta', D) \vdash^{*} C} {{(\Gamma'', A, B; \Delta', D) \vdash^{*} C}}}\quad \quad {}}

These can be transformed into derivations with cuts of cut-height $n+m$:
\vspace{0.3cm}

\hspace{-1.3cm}
  \quad
{\footnotesize
\infer[\scriptstyle\wedge L^{a}]{(\Gamma, \Gamma'', A \wedge B; \Delta, \Delta') \vdash^{*} C} 
{\infer[\scriptstyle Cut^{a}]{(\Gamma, \Gamma'', A, B; \Delta, \Delta') \vdash^{*} C}
{{(\Gamma; \Delta) \vdash^{+} D} 
\quad \quad {(\Gamma'', A, B, D; \Delta') \vdash^{*} C}}}
\quad \quad
\infer[\scriptstyle\wedge L^{a}]{(\Gamma, \Gamma'', A \wedge B; \Delta, \Delta') \vdash^{*} C} 
{\infer[\scriptstyle Cut^{c}]{(\Gamma, \Gamma'', A, B; \Delta, \Delta') \vdash^{*} C}
{{(\Gamma; \Delta) \vdash^{-} D} 
\quad \quad {(\Gamma'', A, B; \Delta', D) \vdash^{*} C}}}}

\item[-4.2-] $ \wedge L^{c} $ is the last rule used to derive the right premise with $\Delta' = \Delta'', A \wedge B$. The derivations with cuts of cut-height $n+max (m,k)+1$ are
	\vspace{0.3cm}

\hspace{-1.3cm}
  \quad 
{\small\infer[\scriptstyle Cut^{a}]{(\Gamma, \Gamma'; \Delta, \Delta'', A \wedge B) \vdash^{*} C}{\quad \quad \quad \quad{(\Gamma; \Delta) \vdash^{+} D}\quad \infer[\scriptstyle\wedge L^{c}]{(\Gamma', D; \Delta'', A \wedge B) \vdash^{*} C} {{(\Gamma', D; \Delta'', A) \vdash^{*} C} \quad {(\Gamma', D; \Delta'', B) \vdash^{*} C}}}   
\vspace{0.5cm}

\infer[\scriptstyle Cut^{c}]{(\Gamma, \Gamma'; \Delta, \Delta'', A \wedge B) \vdash^{*} C}{\quad \quad \quad \quad{(\Gamma; \Delta) \vdash^{-} D}\quad \infer[\scriptstyle\wedge L^{c}]{(\Gamma'; \Delta'', A \wedge B, D) \vdash^{*} C} {{(\Gamma'; \Delta'', A, D) \vdash^{*} C} \quad {(\Gamma'; \Delta'', B, D) \vdash^{*} C}}}}

These can be transformed into derivations each with two cuts of cut-height $n+m$ and $n+k$, respectively:
\vspace{0.3cm}

\hspace{-1.3cm}
  \quad
{\small\infer[\scriptstyle\wedge L^{c}]{(\Gamma, \Gamma'; \Delta, \Delta'', A \wedge B) \vdash^{*} C}{\infer[\scriptstyle Cut^{a}]{(\Gamma, \Gamma'; \Delta, \Delta'', A) \vdash^{*} C} {{(\Gamma; \Delta) \vdash^{+} D} \quad {(\Gamma', D; \Delta'', A) \vdash^{*} C}} \quad \quad {\infer[\scriptstyle Cut^{a}]{(\Gamma, \Gamma'; \Delta, \Delta'', B) \vdash^{*} C} {{(\Gamma; \Delta) \vdash^{+} D} \quad {(\Gamma', D; \Delta'', B) \vdash^{*} C}}}}   
\vspace{0.5cm}

\infer[\scriptstyle\wedge L^{c}]{(\Gamma, \Gamma'; \Delta, \Delta'', A \wedge B) \vdash^{*} C}{\infer[\scriptstyle Cut^{c}]{(\Gamma, \Gamma'; \Delta, \Delta'', A) \vdash^{*} C} {{(\Gamma; \Delta) \vdash^{-} D} \quad {(\Gamma'; \Delta'', A, D) \vdash^{*} C}} \quad \quad {\infer[\scriptstyle Cut^{c}]{(\Gamma, \Gamma'; \Delta, \Delta'', B) \vdash^{*} C} {{(\Gamma; \Delta) \vdash^{-} D} \quad {(\Gamma'; \Delta'', B, D) \vdash^{*} C}}}}}

\item[-4.3-] $ \vee L^{a} $ is the last rule used to derive the right premise with $\Gamma' = \Gamma'', A \vee B$. The derivations with cuts of cut-height $n+max (m,k)+1$ are
	\vspace{0.3cm}

\hspace{-1.3cm}
  \quad
{\small\infer[\scriptstyle Cut^{a}]{(\Gamma, \Gamma'', A \vee B; \Delta, \Delta') \vdash^{*} C}{\quad \quad \quad \quad{(\Gamma; \Delta) \vdash^{+} D}\quad \infer[\scriptstyle\vee L^{a}]{(\Gamma'', A \vee B, D; \Delta') \vdash^{*} C} {{(\Gamma'', A, D; \Delta') \vdash^{*} C} \quad {(\Gamma'', B, D; \Delta') \vdash^{*} C}}}   
\vspace{0.5cm}

\infer[\scriptstyle Cut^{c}]{(\Gamma, \Gamma'', A \vee B; \Delta, \Delta') \vdash^{*} C}{\quad \quad \quad \quad{(\Gamma; \Delta) \vdash^{-} D}\quad \infer[\scriptstyle\vee L^{a}]{(\Gamma'', A \vee B; \Delta', D) \vdash^{*} C} {{(\Gamma'', A; \Delta', D) \vdash^{*} C} \quad {(\Gamma'', B; \Delta', D) \vdash^{*} C}}}}

These can be transformed into derivations each with two cuts of cut-height $n+m$ and $n+k$, respectively:
\vspace{0.3cm}

\hspace{-1.3cm}
  \quad
{\small\infer[\scriptstyle\vee L^{a}]{(\Gamma, \Gamma'', A \vee B; \Delta, \Delta') \vdash^{*} C}{\infer[\scriptstyle Cut^{a}]{(\Gamma, \Gamma'', A; \Delta, \Delta') \vdash^{*} C} {{(\Gamma; \Delta) \vdash^{+} D} \quad {(\Gamma'', A, D; \Delta') \vdash^{*} C}} \quad \quad {\infer[\scriptstyle Cut^{a}]{(\Gamma, \Gamma'', B; \Delta, \Delta') \vdash^{*} C} {{(\Gamma; \Delta) \vdash^{+} D} \quad {(\Gamma'', B, D; \Delta') \vdash^{*} C}}}}   
\vspace{0.5cm}

\infer[\scriptstyle\vee L^{a}]{(\Gamma, \Gamma'', A \vee B; \Delta, \Delta') \vdash^{*} C}{\infer[\scriptstyle Cut^{c}]{(\Gamma, \Gamma'', A; \Delta, \Delta') \vdash^{*} C} {{(\Gamma; \Delta) \vdash^{-} D} \quad {(\Gamma'', A; \Delta', D) \vdash^{*} C}} \quad \quad {\infer[\scriptstyle Cut^{c}]{(\Gamma, \Gamma'', B; \Delta, \Delta') \vdash^{*} C} {{(\Gamma; \Delta) \vdash^{-} D} \quad {(\Gamma'', B; \Delta', D) \vdash^{*} C}}}}}

\item[-4.4-] $ \vee L^{c} $ is the last rule used to derive the right premise with $\Delta' = \Delta'', A \vee B$. The derivations with cuts of cut-height $n+m+1$ are
	\vspace{0.3cm}

\hspace{-1.3cm}
  \quad
{\footnotesize\infer[\scriptstyle Cut^{a}]{(\Gamma, \Gamma'; \Delta, \Delta'', A \vee B) \vdash^{*} C}{{(\Gamma; \Delta) \vdash^{+} D}\quad \quad {}\infer[\scriptstyle\vee L^{c}]{(\Gamma', D; \Delta'', A \vee B) \vdash^{*} C} {{(\Gamma', D; \Delta'', A, B) \vdash^{*} C}}}
\quad \quad    
\infer[\scriptstyle Cut^{c}]{(\Gamma, \Gamma'; \Delta, \Delta'', A \vee B) \vdash^{*} C}{{(\Gamma; \Delta) \vdash^{-} D}\quad \quad {}\infer[\scriptstyle\vee L^{c}]{(\Gamma'; \Delta'', A \vee B, D) \vdash^{*} C} {{(\Gamma'; \Delta'', A, B, D) \vdash^{*} C}}}\quad \quad {}}

These can be transformed into derivations with cuts of cut-height $n+m$:
\vspace{0.3cm}

\hspace{-1.3cm}
  \quad
{\footnotesize
\infer[\scriptstyle\vee L^{c}]{(\Gamma, \Gamma'; \Delta, \Delta'' A \vee B) \vdash^{*} C} 
{\infer[\scriptstyle Cut^{a}]{(\Gamma, \Gamma'; \Delta, \Delta'', A, B) \vdash^{*} C}
{{(\Gamma; \Delta) \vdash^{+} D} 
\quad \quad {(\Gamma', D; \Delta'', A, B) \vdash^{*} C}}}
\quad \quad
\infer[\scriptstyle\vee L^{c}]{(\Gamma, \Gamma'; \Delta, \Delta'', A \vee B) \vdash^{*} C} 
{\infer[\scriptstyle Cut^{c}]{(\Gamma, \Gamma'; \Delta, \Delta'', A, B) \vdash^{*} C}
{{(\Gamma; \Delta) \vdash^{-} D} 
\quad \quad {(\Gamma'; \Delta'', A, B, D) \vdash^{*} C}}}}

\item[-4.5-] $ \rightarrow L^{a} $ is the last rule used to derive the right premise with $\Gamma' = \Gamma'', A \rightarrow B$. The derivations with cuts of cut-height $n+max (m,k)+1$ are
	\vspace{0.3cm}

\hspace{-1.3cm}
  \quad
{\small\infer[\scriptstyle Cut^{a}]{(\Gamma, \Gamma'', A \rightarrow B; \Delta, \Delta') \vdash^{*} C}{\quad \quad \quad \quad \quad{(\Gamma; \Delta) \vdash^{+} D}\quad \infer[\scriptstyle\rightarrow L^{a}]{(\Gamma'', A \rightarrow B, D; \Delta') \vdash^{*} C} {{(\Gamma'', A \rightarrow B, D; \Delta') \vdash^{+} A} \quad {(\Gamma'', B, D; \Delta') \vdash^{*} C}}}  
\vspace{0.5cm}
 
\infer[\scriptstyle Cut^{c}]{(\Gamma, \Gamma'', A \rightarrow B; \Delta, \Delta') \vdash^{*} C}{\quad \quad \quad \quad \quad{(\Gamma; \Delta) \vdash^{-} D}\quad \infer[\scriptstyle\rightarrow L^{a}]{(\Gamma'', A \rightarrow B; \Delta', D) \vdash^{*} C} {{(\Gamma'', A \rightarrow B; \Delta', D) \vdash^{+} A} \quad {(\Gamma'', B; \Delta', D) \vdash^{*} C}}}}

These can be transformed into derivations each with two cuts of cut-height $n+m$ and $n+k$, respectively:
\vspace{0.3cm}

\hspace{-1.3cm}
  \quad
{\small\infer[\scriptstyle\rightarrow L^{a}]{(\Gamma, \Gamma'', A \rightarrow B; \Delta, \Delta') \vdash^{*} C}{\infer[\scriptstyle Cut^{a}]{(\Gamma, \Gamma'', A \rightarrow B; \Delta, \Delta') \vdash^{+} A} {{(\Gamma; \Delta) \vdash^{+} D} \quad {(\Gamma'', A \rightarrow B, D; \Delta') \vdash^{+} A}} \quad \quad {\infer[\scriptstyle Cut^{a}]{(\Gamma, \Gamma'', B; \Delta, \Delta') \vdash^{*} C} {{(\Gamma; \Delta) \vdash^{+} D} \quad {(\Gamma'', B, D; \Delta') \vdash^{*} C}}}}   
\vspace{0.5cm}

\infer[\scriptstyle\rightarrow L^{a}]{(\Gamma, \Gamma'', A \rightarrow B; \Delta, \Delta') \vdash^{*} C}{\infer[\scriptstyle Cut^{c}]{(\Gamma, \Gamma'', A \rightarrow B; \Delta, \Delta') \vdash^{+} A} {{(\Gamma; \Delta) \vdash^{-} D} \quad {(\Gamma'', A \rightarrow B; \Delta', D) \vdash^{+} A}} \quad \quad {\infer[\scriptstyle Cut^{c}]{(\Gamma, \Gamma'', B; \Delta, \Delta') \vdash^{*} C} {{(\Gamma; \Delta) \vdash^{-} D} \quad {(\Gamma'', B; \Delta', D) \vdash^{*} C}}}}}

\item[-4.6-] $ \rightarrow L^{c}$ is the last rule used to derive the right premise with $\Delta' = \Delta'', A \rightarrow B$. The derivations with cuts of cut-height $n+m+1$ are
	\vspace{0.3cm}

\hspace{-1.3cm}
  \quad
{\footnotesize\infer[\scriptstyle Cut^{a}]{(\Gamma, \Gamma'; \Delta, \Delta'', A \rightarrow B) \vdash^{*} C}{{(\Gamma; \Delta) \vdash^{+} D}\quad \quad {}\infer[\scriptstyle\rightarrow L^{c}]{(\Gamma', D; \Delta'', A \rightarrow B) \vdash^{*} C} {{(\Gamma', A, D; \Delta'', B) \vdash^{*} C}}}
\quad     
\infer[\scriptstyle Cut^{c}]{(\Gamma, \Gamma'; \Delta, \Delta'', A \rightarrow B) \vdash^{*} C}{{(\Gamma; \Delta) \vdash^{-} D}\quad \quad {}\infer[\scriptstyle\rightarrow L^{c}]{(\Gamma'; \Delta'', A \rightarrow B, D) \vdash^{*} C} {{(\Gamma', A; \Delta'', B, D) \vdash^{*} C}}}\quad \quad {}}

These can be transformed into derivations with cuts of cut-height $n+m$:
\vspace{0.3cm}

\hspace{-1.3cm}
  \quad
{\footnotesize
\infer[\scriptstyle\rightarrow L^{c}]{(\Gamma, \Gamma'; \Delta, \Delta'', A \rightarrow B) \vdash^{*} C} 
{\infer[\scriptstyle Cut^{a}]{(\Gamma, \Gamma', A; \Delta, \Delta'', B) \vdash^{*} C}
{{(\Gamma; \Delta) \vdash^{+} D} 
\quad \quad {(\Gamma', A, D; \Delta'', B) \vdash^{*} C}}}
\quad \quad
\infer[\scriptstyle\rightarrow L^{c}]{(\Gamma, \Gamma'; \Delta, \Delta'', A \rightarrow B) \vdash^{*} C} 
{\infer[\scriptstyle Cut^{c}]{(\Gamma, \Gamma', A; \Delta, \Delta'', B) \vdash^{*} C}
{{(\Gamma; \Delta) \vdash^{-} D} 
\quad \quad {(\Gamma', A; \Delta'',B, D) \vdash^{*} C}}}}

\item[-4.7-] $ \Yleft L^{a}$ is the last rule used to derive the right premise with $\Gamma' = \Gamma'', A \Yleft B$. The derivations with cuts of cut-height $n+m+1$ are
	\vspace{0.3cm}

\hspace{-1.3cm}
  \quad
{\footnotesize\infer[\scriptstyle Cut^{a}]{(\Gamma, \Gamma'', A \Yleft B; \Delta, \Delta') \vdash^{*} C}{{(\Gamma; \Delta) \vdash^{+} D}\quad \quad {}\infer[\scriptstyle\Yleft L^{a}]{(\Gamma'', A \Yleft B, D; \Delta') \vdash^{*} C} {{(\Gamma'', A, D; \Delta', B) \vdash^{*} C}}}
\quad     
\infer[\scriptstyle Cut^{c}]{(\Gamma, \Gamma'', A \Yleft B; \Delta, \Delta') \vdash^{*} C}{{(\Gamma; \Delta) \vdash^{-} D}\quad \quad {}\infer[\scriptstyle\Yleft L^{a}]{(\Gamma'', A \Yleft B; \Delta', D) \vdash^{*} C} {{(\Gamma'', A; \Delta', B, D) \vdash^{*} C}}}\quad \quad {}}

These can be transformed into derivations with cuts of cut-height $n+m$:
\vspace{0.3cm}

\hspace{-1.3cm}
  \quad
{\footnotesize
\infer[\scriptstyle\Yleft L^{a}]{(\Gamma, \Gamma'', A \Yleft B; \Delta, \Delta') \vdash^{*} C} 
{\infer[\scriptstyle Cut^{a}]{(\Gamma, \Gamma'', A; \Delta, \Delta', B) \vdash^{*} C}
{{(\Gamma; \Delta) \vdash^{+} D} 
\quad \quad {(\Gamma'', A, D; \Delta', B) \vdash^{*} C}}}
\quad \quad
\infer[\scriptstyle\Yleft L^{a}]{(\Gamma, \Gamma'', A \Yleft B; \Delta, \Delta') \vdash^{*} C} 
{\infer[\scriptstyle Cut^{c}]{(\Gamma, \Gamma'', A; \Delta, \Delta', B) \vdash^{*} C}
{{(\Gamma; \Delta) \vdash^{-} D} 
\quad \quad {(\Gamma'', A; \Delta', B, D) \vdash^{*} C}}}}

\item[-4.8-] $ \Yleft L^{c} $ is the last rule used to derive the right premise with $\Delta' = \Delta'', A \Yleft B$. The derivations with cuts of cut-height $n+max (m,k)+1$ are
	\vspace{0.3cm}

\hspace{-1.3cm}
  \quad
{\small\infer[\scriptstyle Cut^{a}]{(\Gamma, \Gamma'; \Delta, \Delta'', A \Yleft B) \vdash^{*} C}{\quad \quad \quad \quad \quad{(\Gamma; \Delta) \vdash^{+} D}\quad \infer[\scriptstyle\Yleft L^{c}]{(\Gamma', D; \Delta'', A \Yleft B) \vdash^{*} C} {{(\Gamma', D; \Delta'', A \Yleft B) \vdash^{-} B} \quad {(\Gamma', D; \Delta'', A) \vdash^{*} C}}}  
\vspace{0.5cm}
 
\infer[\scriptstyle Cut^{c}]{(\Gamma, \Gamma'; \Delta, \Delta'', A \Yleft B) \vdash^{*} C}{\quad \quad \quad \quad \quad{(\Gamma; \Delta) \vdash^{-} D}\quad \infer[\scriptstyle\Yleft L^{c}]{(\Gamma'; \Delta'', A \Yleft B, D) \vdash^{*} C} {{(\Gamma'; \Delta'', A \Yleft B, D) \vdash^{-} B} \quad {(\Gamma'; \Delta'', A, D) \vdash^{*} C}}}}

These can be transformed into derivations each with two cuts of cut-height $n+m$ and $n+k$, respectively:
\vspace{0.3cm}

\hspace{-1.3cm}
  \quad
{\small\infer[\scriptstyle\Yleft L^{c}]{(\Gamma, \Gamma'; \Delta, \Delta'', A \Yleft B) \vdash^{*} C}{\infer[\scriptstyle Cut^{a}]{(\Gamma, \Gamma'; \Delta, \Delta'', A \Yleft B) \vdash^{-} B} {{(\Gamma; \Delta) \vdash^{+} D} \quad {(\Gamma', D; \Delta'', A \Yleft B) \vdash^{-} B}} \quad \quad {\infer[\scriptstyle Cut^{a}]{(\Gamma, \Gamma'; \Delta, \Delta'', A) \vdash^{*} C} {{(\Gamma; \Delta) \vdash^{+} D} \quad {(\Gamma', D; \Delta'', A) \vdash^{*} C}}}}   
\vspace{0.5cm}

\infer[\scriptstyle\Yleft L^{c}]{(\Gamma, \Gamma'; \Delta, \Delta'', A \Yleft B) \vdash^{*} C}{\infer[\scriptstyle Cut^{c}]{(\Gamma, \Gamma'; \Delta, \Delta'', A \Yleft B) \vdash^{-} B} {{(\Gamma; \Delta) \vdash^{-} D} \quad {(\Gamma'; \Delta'', A \Yleft B, D) \vdash^{-} B}} \quad \quad {\infer[\scriptstyle Cut^{c}]{(\Gamma, \Gamma'; \Delta, \Delta'', A) \vdash^{*} C} {{(\Gamma; \Delta) \vdash^{-} D} \quad {(\Gamma'; \Delta'', A, D) \vdash^{*} C}}}}}

\item[-4.9-] $ \wedge R^{+} $ is the last rule used to derive the right premise with $C = A \wedge B$. The derivations with cuts of cut-height $n+max (m,k)+1$ are
	\vspace{0.3cm}

\hspace{-2cm}
  \quad
{\scriptsize\infer[\scriptstyle Cut^{a}]{(\Gamma, \Gamma'; \Delta, \Delta') \vdash^{+} A \wedge B}{\quad \quad \quad {(\Gamma; \Delta) \vdash^{+} D}\quad \infer[\scriptstyle\wedge R^{+}]{(\Gamma', D; \Delta') \vdash^{+} A \wedge B} {{(\Gamma', D; \Delta') \vdash^{+} A} \quad {(\Gamma', D; \Delta') \vdash^{+} B}}}   
\infer[\scriptstyle Cut^{c}]{(\Gamma, \Gamma'; \Delta, \Delta') \vdash^{+} A \wedge B}{\quad \quad \quad{(\Gamma; \Delta) \vdash^{-} D}\quad \infer[\scriptstyle\wedge R^{+}]{(\Gamma'; \Delta', D) \vdash^{+} A \wedge B} {{(\Gamma'; \Delta', D) \vdash^{+} A} \quad {(\Gamma'; \Delta', D) \vdash^{+} B}}}}

These can be transformed into derivations each with two cuts of cut-height $n+m$ and $n+k$, respectively:
\vspace{0.3cm}

\hspace{-1.3cm}
  \quad
{\small\infer[\scriptstyle\wedge R^{+}]{(\Gamma, \Gamma'; \Delta, \Delta') \vdash^{+} A \wedge B}{\infer[\scriptstyle Cut^{a}]{(\Gamma, \Gamma'; \Delta, \Delta') \vdash^{+} A} {{(\Gamma; \Delta) \vdash^{+} D} \quad {(\Gamma', D; \Delta') \vdash^{+} A}} \quad \quad {\infer[\scriptstyle Cut^{a}]{(\Gamma, \Gamma'; \Delta, \Delta') \vdash^{+} B} {{(\Gamma; \Delta) \vdash^{+} D} \quad {(\Gamma', D; \Delta') \vdash^{+} B}}}}   
\vspace{0.5cm}

\infer[\scriptstyle\wedge R^{+}]{(\Gamma, \Gamma'; \Delta, \Delta') \vdash^{+} A \wedge B}{\infer[\scriptstyle Cut^{c}]{(\Gamma, \Gamma'; \Delta, \Delta') \vdash^{+} A} {{(\Gamma; \Delta) \vdash^{-} D} \quad {(\Gamma'; \Delta', D) \vdash^{+} A}} \quad \quad {\infer[\scriptstyle Cut^{c}]{(\Gamma, \Gamma'; \Delta, \Delta') \vdash^{+} B} {{(\Gamma; \Delta) \vdash^{-} D} \quad {(\Gamma'; \Delta', D) \vdash^{+} B}}}}}

\item[-4.10.1-] $ \wedge R^{-}_{1} $ is the last rule used to derive the right premise with $C = A \wedge B$. The derivations with cuts of cut-height $n+m+1$ are
	\vspace{0.3cm}

\hspace{-1.3cm}
  \quad
{\footnotesize\infer[\scriptstyle Cut^{a}]{(\Gamma, \Gamma'; \Delta, \Delta') \vdash^{-} A \wedge B}{{(\Gamma; \Delta) \vdash^{+} D}\quad \quad {}\infer[\scriptstyle\wedge R^{-}_{1}]{(\Gamma', D; \Delta') \vdash^{-} A \wedge B} {{(\Gamma', D; \Delta') \vdash^{-} A}}}
\quad \quad    
\infer[\scriptstyle Cut^{c}]{(\Gamma, \Gamma'; \Delta, \Delta') \vdash^{-} A \wedge B}{{(\Gamma; \Delta) \vdash^{-} D}\quad \quad {}\infer[\scriptstyle\wedge R^{-}_{1}]{(\Gamma'; \Delta', D) \vdash^{-} A \wedge B} {{(\Gamma'; \Delta', D) \vdash^{-} A}}}\quad \quad {}}

These can be transformed into derivations with cuts of cut-height $n+m$:
\vspace{0.3cm}

\hspace{-1.3cm}
  \quad
{\footnotesize
\infer[\scriptstyle\wedge R^{-}_{1}]{(\Gamma, \Gamma'; \Delta, \Delta') \vdash^{-} A \wedge B} 
{\infer[\scriptstyle Cut^{a}]{(\Gamma, \Gamma'; \Delta, \Delta') \vdash^{-} A}
{{(\Gamma; \Delta) \vdash^{+} D} 
\quad \quad {(\Gamma', D; \Delta') \vdash^{-} A}}}
\quad \quad
\infer[\scriptstyle\wedge R^{-}_{1}]{(\Gamma, \Gamma'; \Delta, \Delta') \vdash^{-} A \wedge B} 
{\infer[\scriptstyle Cut^{c}]{(\Gamma, \Gamma'; \Delta, \Delta') \vdash^{-} A }
{{(\Gamma; \Delta) \vdash^{-} D} 
\quad \quad {(\Gamma'; \Delta', D) \vdash^{-} A}}}}

\item[-4.10.2-] $ \wedge R^{-}_{2} $ is the last rule used to derive the right premise with $C = A \wedge B$. The derivations with cuts of cut-height $n+m+1$ are
	\vspace{0.3cm}

\hspace{-1.3cm}
  \quad
{\footnotesize\infer[\scriptstyle Cut^{a}]{(\Gamma, \Gamma'; \Delta, \Delta') \vdash^{-} A \wedge B}{{(\Gamma; \Delta) \vdash^{+} D}\quad \quad {}\infer[\scriptstyle\wedge R^{-}_{2}]{(\Gamma', D; \Delta') \vdash^{-} A \wedge B} {{(\Gamma', D; \Delta') \vdash^{-} B}}}
\quad \quad    
\infer[\scriptstyle Cut^{c}]{(\Gamma, \Gamma'; \Delta, \Delta') \vdash^{-} A \wedge B}{{(\Gamma; \Delta) \vdash^{-} D}\quad \quad {}\infer[\scriptstyle\wedge R^{-}_{2}]{(\Gamma'; \Delta', D) \vdash^{-} A \wedge B} {{(\Gamma'; \Delta', D) \vdash^{-} B}}}\quad \quad {}}

These can be transformed into derivations with cuts of cut-height $n+m$:
\vspace{0.3cm}

\hspace{-1.3cm}
  \quad
{\footnotesize
\infer[\scriptstyle\wedge R^{-}_{2}]{(\Gamma, \Gamma'; \Delta, \Delta') \vdash^{-} A \wedge B} 
{\infer[\scriptstyle Cut^{a}]{(\Gamma, \Gamma'; \Delta, \Delta') \vdash^{-} B}
{{(\Gamma; \Delta) \vdash^{+} D} 
\quad \quad {(\Gamma', D; \Delta') \vdash^{-} B}}}
\quad \quad
\infer[\scriptstyle\wedge R^{-}_{2}]{(\Gamma, \Gamma'; \Delta, \Delta') \vdash^{-} A \wedge B} 
{\infer[\scriptstyle Cut^{c}]{(\Gamma, \Gamma'; \Delta, \Delta') \vdash^{-} B }
{{(\Gamma; \Delta) \vdash^{-} D} 
\quad \quad {(\Gamma'; \Delta', D) \vdash^{-} B}}}}

\item[-4.11.1-] $ \vee R^{+}_{1} $ is the last rule used to derive the right premise with $C = A \vee B$. The derivations with cuts of cut-height $n+m+1$ are
	\vspace{0.3cm}

\hspace{-1.3cm}
  \quad
{\footnotesize\infer[\scriptstyle Cut^{a}]{(\Gamma, \Gamma'; \Delta, \Delta') \vdash^{+} A \vee B}{{(\Gamma; \Delta) \vdash^{+} D}\quad \quad {}\infer[\scriptstyle\vee R^{+}_{1}]{(\Gamma', D; \Delta') \vdash^{+} A \vee B} {{(\Gamma', D; \Delta') \vdash^{+} A}}}
\quad \quad    
\infer[\scriptstyle Cut^{c}]{(\Gamma, \Gamma'; \Delta, \Delta') \vdash^{+} A \vee B}{{(\Gamma; \Delta) \vdash^{-} D}\quad \quad {}\infer[\scriptstyle\vee R^{+}_{1}]{(\Gamma'; \Delta', D) \vdash^{+} A \vee B} {{(\Gamma'; \Delta', D) \vdash^{+} A}}}\quad \quad {}}

These can be transformed into derivations with cuts of cut-height $n+m$:
\vspace{0.3cm}

\hspace{-1.3cm}
  \quad
{\footnotesize
\infer[\scriptstyle\vee R^{+}_{1}]{(\Gamma, \Gamma'; \Delta, \Delta') \vdash^{+} A \vee B} 
{\infer[\scriptstyle Cut^{a}]{(\Gamma, \Gamma'; \Delta, \Delta') \vdash^{+} A}
{{(\Gamma; \Delta) \vdash^{+} D} 
\quad \quad {(\Gamma', D; \Delta') \vdash^{+} A}}}
\quad \quad
\infer[\scriptstyle\vee R^{+}_{1}]{(\Gamma, \Gamma'; \Delta, \Delta') \vdash^{+} A \vee B} 
{\infer[\scriptstyle Cut^{c}]{(\Gamma, \Gamma'; \Delta, \Delta') \vdash^{+} A }
{{(\Gamma; \Delta) \vdash^{-} D} 
\quad \quad {(\Gamma'; \Delta', D) \vdash^{+} A}}}}

\item[-4.11.2-] $ \vee R^{+}_{2} $ is the last rule used to derive the right premise with $C = A \vee B$. The derivations with cuts of cut-height $n+m+1$ are
	\vspace{0.3cm}

\hspace{-1.3cm}
  \quad
{\footnotesize\infer[\scriptstyle Cut^{a}]{(\Gamma, \Gamma'; \Delta, \Delta') \vdash^{+} A \vee B}{{(\Gamma; \Delta) \vdash^{+} D}\quad \quad {}\infer[\scriptstyle\vee R^{+}_{2}]{(\Gamma', D; \Delta') \vdash^{+} A \vee B} {{(\Gamma', D; \Delta') \vdash^{+} B}}}
\quad \quad    
\infer[\scriptstyle Cut^{c}]{(\Gamma, \Gamma'; \Delta, \Delta') \vdash^{+} A \vee B}{{(\Gamma; \Delta) \vdash^{-} D}\quad \quad {}\infer[\scriptstyle\vee R^{+}_{2}]{(\Gamma'; \Delta', D) \vdash^{+} A \vee B} {{(\Gamma'; \Delta', D) \vdash^{+} B}}}\quad \quad {}}

These can be transformed into derivations with cuts of cut-height $n+m$:
\vspace{0.3cm}

\hspace{-1.3cm}
  \quad
{\footnotesize
\infer[\scriptstyle\vee R^{+}_{2}]{(\Gamma, \Gamma'; \Delta, \Delta') \vdash^{+} A \vee B} 
{\infer[\scriptstyle Cut^{a}]{(\Gamma, \Gamma'; \Delta, \Delta') \vdash^{+} B}
{{(\Gamma; \Delta) \vdash^{+} D} 
\quad \quad {(\Gamma', D; \Delta') \vdash^{+} B}}}
\quad \quad
\infer[\scriptstyle\vee R^{+}_{2}]{(\Gamma, \Gamma'; \Delta, \Delta') \vdash^{+} A \vee B} 
{\infer[\scriptstyle Cut^{c}]{(\Gamma, \Gamma'; \Delta, \Delta') \vdash^{+} B }
{{(\Gamma; \Delta) \vdash^{-} D} 
\quad \quad {(\Gamma'; \Delta', D) \vdash^{+} B}}}}

\item[-4.12-] $ \vee R^{-} $ is the last rule used to derive the right premise with $C = A \vee B$. The derivations with cuts of cut-height $n+max (m,k)+1$ are
	\vspace{0.3cm}

\hspace{-2cm}
  \quad
{\scriptsize\infer[\scriptstyle Cut^{a}]{(\Gamma, \Gamma'; \Delta, \Delta') \vdash^{-} A \vee B}{\quad \quad \quad {(\Gamma; \Delta) \vdash^{+} D}\quad \infer[\scriptstyle\vee R^{-}]{(\Gamma', D; \Delta') \vdash^{-} A \vee B} {{(\Gamma', D; \Delta') \vdash^{-} A} \quad {(\Gamma', D; \Delta') \vdash^{-} B}}}  

\infer[\scriptstyle Cut^{c}]{(\Gamma, \Gamma'; \Delta, \Delta') \vdash^{-} A \vee B}{\quad \quad \quad{(\Gamma; \Delta) \vdash^{-} D}\quad \infer[\scriptstyle\vee R^{-}]{(\Gamma'; \Delta', D) \vdash^{-} A \vee B} {{(\Gamma'; \Delta', D) \vdash^{-} A} \quad {(\Gamma'; \Delta', D) \vdash^{-} B}}}}

These can be transformed into derivations each with two cuts of cut-height $n+m$ and $n+k$, respectively:
\vspace{0.3cm}

\hspace{-1cm}
  \quad
{\small\infer[\scriptstyle\vee R^{-}]{(\Gamma, \Gamma'; \Delta, \Delta') \vdash^{-} A \vee B}{\infer[\scriptstyle Cut^{a}]{(\Gamma, \Gamma'; \Delta, \Delta') \vdash^{-} A} {{(\Gamma; \Delta) \vdash^{+} D} \quad {(\Gamma', D; \Delta') \vdash^{-} A}} \quad \quad {\infer[\scriptstyle Cut^{a}]{(\Gamma, \Gamma'; \Delta, \Delta') \vdash^{-} B} {{(\Gamma; \Delta) \vdash^{+} D} \quad {(\Gamma', D; \Delta') \vdash^{-} B}}}}   
\vspace{0.5cm}

\hspace{-0.5cm} 
\infer[\scriptstyle\vee R^{-}]{(\Gamma, \Gamma'; \Delta, \Delta') \vdash^{-} A \vee B}{\infer[\scriptstyle Cut^{c}]{(\Gamma, \Gamma'; \Delta, \Delta') \vdash^{-} A} {{(\Gamma; \Delta) \vdash^{-} D} \quad {(\Gamma'; \Delta', D) \vdash^{-} A}} \quad \quad {\infer[\scriptstyle Cut^{c}]{(\Gamma, \Gamma'; \Delta, \Delta') \vdash^{-} B} {{(\Gamma; \Delta) \vdash^{-} D} \quad {(\Gamma'; \Delta', D) \vdash^{-} B}}}}}

\item[-4.13-] $ \rightarrow R^{+} $ is the last rule used to derive the right premise with $C = A \rightarrow B$. The derivations with cuts of cut-height $n+m+1$ are
	\vspace{0.3cm}

\hspace{-1.3cm}
  \quad
{\footnotesize\infer[\scriptstyle Cut^{a}]{(\Gamma, \Gamma'; \Delta, \Delta') \vdash^{+} A \rightarrow B}{{(\Gamma; \Delta) \vdash^{+} D}\quad \quad {}\infer[\scriptstyle\rightarrow R^{+}]{(\Gamma', D; \Delta') \vdash^{+} A \rightarrow B} {{(\Gamma', A, D; \Delta') \vdash^{+} B}}}
\quad \quad    
\infer[\scriptstyle Cut^{c}]{(\Gamma, \Gamma'; \Delta, \Delta') \vdash^{+} A \rightarrow B}{{(\Gamma; \Delta) \vdash^{-} D}\quad \quad {}\infer[\scriptstyle\rightarrow R^{+}]{(\Gamma'; \Delta', D) \vdash^{+} A \rightarrow B} {{(\Gamma', A; \Delta', D) \vdash^{+} B}}}\quad \quad {}}

These can be transformed into derivations with cuts of cut-height $n+m$:
\vspace{0.3cm}

\hspace{-1.3cm}
  \quad
{\footnotesize
\infer[\scriptstyle\rightarrow R^{+}]{(\Gamma, \Gamma'; \Delta, \Delta') \vdash^{+} A \rightarrow B} 
{\infer[\scriptstyle Cut^{a}]{(\Gamma, \Gamma', A; \Delta, \Delta') \vdash^{+} B}
{{(\Gamma; \Delta) \vdash^{+} D} 
\quad \quad {(\Gamma', A, D; \Delta') \vdash^{+} B}}}
\quad \quad
\infer[\scriptstyle\rightarrow R^{+}]{(\Gamma, \Gamma'; \Delta, \Delta') \vdash^{+} A \rightarrow B} 
{\infer[\scriptstyle Cut^{c}]{(\Gamma, \Gamma', A; \Delta, \Delta') \vdash^{+} B }
{{(\Gamma; \Delta) \vdash^{-} D} 
\quad \quad {(\Gamma', A; \Delta', D) \vdash^{+} B}}}}

\item[-4.14-] $ \rightarrow R^{-} $ is the last rule used to derive the right premise with $C = A \rightarrow B$. The derivations with cuts of cut-height $n+max (m,k)+1$ are
	\vspace{0.3cm}

\hspace{-2cm}
  \quad
{\scriptsize\infer[\scriptstyle Cut^{a}]{(\Gamma, \Gamma'; \Delta, \Delta') \vdash^{-} A \rightarrow B}{\quad \quad \quad {(\Gamma; \Delta) \vdash^{+} D}\quad \infer[\scriptstyle\rightarrow R^{-}]{(\Gamma', D; \Delta') \vdash^{-} A \rightarrow B} {{(\Gamma', D; \Delta') \vdash^{+} A} \quad {(\Gamma', D; \Delta') \vdash^{-} B}}}   
\infer[\scriptstyle Cut^{c}]{(\Gamma, \Gamma'; \Delta, \Delta') \vdash^{-} A \rightarrow B}{\quad \quad \quad{(\Gamma; \Delta) \vdash^{-} D}\quad \infer[\scriptstyle\rightarrow R^{-}]{(\Gamma'; \Delta', D) \vdash^{-} A \rightarrow B} {{(\Gamma'; \Delta', D) \vdash^{+} A} \quad {(\Gamma'; \Delta', D) \vdash^{-} B}}}}

These can be transformed into derivations each with two cuts of cut-height $n+m$ and $n+k$, respectively:
\vspace{0.3cm}

\hspace{-1cm}
  \quad
{\small\infer[\scriptstyle\rightarrow R^{-}]{(\Gamma, \Gamma'; \Delta, \Delta') \vdash^{-} A \rightarrow B}{\infer[\scriptstyle Cut^{a}]{(\Gamma, \Gamma'; \Delta, \Delta') \vdash^{+} A} {{(\Gamma; \Delta) \vdash^{+} D} \quad {(\Gamma', D; \Delta') \vdash^{+} A}} \quad \quad {\infer[\scriptstyle Cut^{a}]{(\Gamma, \Gamma'; \Delta, \Delta') \vdash^{-} B} {{(\Gamma; \Delta) \vdash^{+} D} \quad {(\Gamma', D; \Delta') \vdash^{-} B}}}}   
\vspace{0.5cm}

\hspace{-0.5cm} 
\infer[\scriptstyle\rightarrow R^{-}]{(\Gamma, \Gamma'; \Delta, \Delta') \vdash^{-} A \rightarrow B}{\infer[\scriptstyle Cut^{c}]{(\Gamma, \Gamma'; \Delta, \Delta') \vdash^{+} A} {{(\Gamma; \Delta) \vdash^{-} D} \quad {(\Gamma'; \Delta', D) \vdash^{+} A}} \quad \quad {\infer[\scriptstyle Cut^{c}]{(\Gamma, \Gamma'; \Delta, \Delta') \vdash^{-} B} {{(\Gamma; \Delta) \vdash^{-} D} \quad {(\Gamma'; \Delta', D) \vdash^{-} B}}}}}

\item[-4.15-] $ \Yleft R^{+} $ is the last rule used to derive the right premise with $C = A \Yleft B$. The derivations with cuts of cut-height $n+max (m,k)+1$ are
	\vspace{0.3cm}

\hspace{-2cm}
  \quad
{\scriptsize\infer[\scriptstyle Cut^{a}]{(\Gamma, \Gamma'; \Delta, \Delta') \vdash^{+} A \Yleft B}{\quad \quad \quad {(\Gamma; \Delta) \vdash^{+} D}\quad \infer[\scriptstyle\Yleft R^{+}]{(\Gamma', D; \Delta') \vdash^{+} A \Yleft B} {{(\Gamma', D; \Delta') \vdash^{+} A} \quad {(\Gamma', D; \Delta') \vdash^{-} B}}}   
\infer[\scriptstyle Cut^{c}]{(\Gamma, \Gamma'; \Delta, \Delta') \vdash^{+} A \Yleft B}{\quad \quad \quad{(\Gamma; \Delta) \vdash^{-} D}\quad \infer[\scriptstyle\Yleft R^{+}]{(\Gamma'; \Delta', D) \vdash^{+} A \Yleft B} {{(\Gamma'; \Delta', D) \vdash^{+} A} \quad {(\Gamma'; \Delta', D) \vdash^{-} B}}}}

These can be transformed into derivations each with two cuts of cut-height $n+m$ and $n+k$, respectively:
\vspace{0.3cm}

\hspace{-1cm}
  \quad
{\small\infer[\scriptstyle\Yleft R^{+}]{(\Gamma, \Gamma'; \Delta, \Delta') \vdash^{+} A \Yleft B}{\infer[\scriptstyle Cut^{a}]{(\Gamma, \Gamma'; \Delta, \Delta') \vdash^{+} A} {{(\Gamma; \Delta) \vdash^{+} D} \quad {(\Gamma', D; \Delta') \vdash^{+} A}} \quad \quad {\infer[\scriptstyle Cut^{a}]{(\Gamma, \Gamma'; \Delta, \Delta') \vdash^{-} B} {{(\Gamma; \Delta) \vdash^{+} D} \quad {(\Gamma', D; \Delta') \vdash^{-} B}}}}   
\vspace{0.5cm}

\hspace{-0.5cm} 
\infer[\scriptstyle\Yleft R^{+}]{(\Gamma, \Gamma'; \Delta, \Delta') \vdash^{+} A \Yleft B}{\infer[\scriptstyle Cut^{c}]{(\Gamma, \Gamma'; \Delta, \Delta') \vdash^{+} A} {{(\Gamma; \Delta) \vdash^{-} D} \quad {(\Gamma'; \Delta', D) \vdash^{+} A}} \quad \quad {\infer[\scriptstyle Cut^{c}]{(\Gamma, \Gamma'; \Delta, \Delta') \vdash^{-} B} {{(\Gamma; \Delta) \vdash^{-} D} \quad {(\Gamma'; \Delta', D) \vdash^{-} B}}}}}

\item[-4.16-] $ \Yleft R^{-} $ is the last rule used to derive the right premise with $C = A \Yleft B$. The derivations with cuts of cut-height $n+m+1$ are
	\vspace{0.3cm}

\hspace{-1.3cm}
  \quad
{\footnotesize\infer[\scriptstyle Cut^{a}]{(\Gamma, \Gamma'; \Delta, \Delta') \vdash^{-} A \Yleft B}{{(\Gamma; \Delta) \vdash^{+} D}\quad \quad {}\infer[\scriptstyle\Yleft R^{-}]{(\Gamma', D; \Delta') \vdash^{-} A \Yleft B} {{(\Gamma', D; \Delta', B) \vdash^{-} A}}}
\quad \quad    
\infer[\scriptstyle Cut^{c}]{(\Gamma, \Gamma'; \Delta, \Delta') \vdash^{-} A \Yleft B}{{(\Gamma; \Delta) \vdash^{-} D}\quad \quad {}\infer[\scriptstyle\Yleft R^{-}]{(\Gamma'; \Delta', D) \vdash^{-} A \Yleft B} {{(\Gamma'; \Delta', B, D) \vdash^{-} A}}}\quad \quad {}}

These can be transformed into derivations with cuts of cut-height $n+m$:
\vspace{0.3cm}

\hspace{-1.3cm}
  \quad
{\footnotesize
\infer[\scriptstyle\Yleft R^{-}]{(\Gamma, \Gamma'; \Delta, \Delta') \vdash^{-} A \Yleft B} 
{\infer[\scriptstyle Cut^{a}]{(\Gamma, \Gamma'; \Delta, \Delta', B) \vdash^{-} A}
{{(\Gamma; \Delta) \vdash^{+} D} 
\quad \quad {(\Gamma', D; \Delta', B) \vdash^{-} A}}}
\quad \quad
\infer[\scriptstyle\Yleft R^{-}]{(\Gamma, \Gamma'; \Delta, \Delta') \vdash^{-} A \Yleft B} 
{\infer[\scriptstyle Cut^{c}]{(\Gamma, \Gamma'; \Delta, \Delta', B) \vdash^{-} A }
{{(\Gamma; \Delta) \vdash^{-} D} 
\quad \quad {(\Gamma'; \Delta', B, D) \vdash^{-} A}}}}
\end{itemize}

It is shown that cut-height is reduced in all cases.

\subsection*{-5- Cut formula \textit{D} principal in both premises}

For each cut rule four cases can be distinguished. Here, it can be shown for each case that the derivations can be transformed into ones in which the occurrences of cut have a reduced cut-height or the cut formula has a lower weight (or both).

\begin{itemize}
	\item[-5.1-] $D = A \wedge B$. The derivation for $Cut^{a}$ with a cut of cut-height $max (n,m)+1+k+1$ is
	\vspace{0.3cm}

  \quad
{\small\infer[\scriptstyle Cut^{a}]{(\Gamma, \Gamma'; \Delta, \Delta') \vdash^{*} C}{\infer[\scriptstyle\wedge R^{+}]{(\Gamma; \Delta) \vdash^{+} A \wedge B} {{(\Gamma; \Delta) \vdash^{+} A} \quad {(\Gamma; \Delta) \vdash^{+} B}} \quad \quad 
{\infer[\scriptstyle\wedge L^{a}]{(\Gamma', A \wedge B; \Delta') \vdash^{*} C} {{(\Gamma', A, B; \Delta') \vdash^{*} C}}}}}   

and can be transformed into a derivation with two cuts of cut-height (from top to bottom) $n+k$ and $m+max(n,k)+1$:
\vspace{0.3cm}

  \quad
{\small
\infer[\scriptstyle C_{a/c}]{(\Gamma, \Gamma'; \Delta, \Delta') \vdash^{*} C} 
{\infer[\scriptstyle Cut^{a}]{(\Gamma, \Gamma, \Gamma'; \Delta, \Delta, \Delta') \vdash^{*} C}
{\quad \quad \quad {(\Gamma; \Delta) \vdash^{+} B} 
\quad \quad {\infer[\scriptstyle Cut^{a}]{(\Gamma, \Gamma', B; \Delta, \Delta') \vdash^{*} C}{{(\Gamma; \Delta) \vdash^{+} A} \quad \quad {(\Gamma', A, B; \Delta') \vdash^{*} C}}}}}
}

Note that in both cases the weight of the cut formula is reduced.
The upper cut is also reduced in height, while with the lower cut we have a case where cut-height is not necessarily reduced. 

The possible derivations for $Cut^{c}$ with a cut of cut-height $n + 1+max(m, k)+1$ are
	\vspace{0.3cm}

  \quad
{\small
\infer[\scriptstyle Cut^{c}]{(\Gamma, \Gamma'; \Delta, \Delta') \vdash^{*} C}{\infer[\scriptstyle\wedge R^{-}_{1}]{\quad \quad \quad (\Gamma; \Delta) \vdash^{-} A \wedge B} {\quad{(\Gamma; \Delta) \vdash^{-} A}} \quad \quad {\infer[\scriptstyle\wedge L^{c}]{(\Gamma';\Delta', A \wedge B) \vdash^{*} C} {{(\Gamma'; \Delta', A) \vdash^{*} C} \quad {(\Gamma'; \Delta', B) \vdash^{*} C}}}}

or

\hspace{-0.1cm} 
\infer[\scriptstyle Cut^{c}]{(\Gamma, \Gamma'; \Delta, \Delta') \vdash^{*} C}{\infer[\scriptstyle\wedge R^{-}_{2}]{\quad \quad \quad (\Gamma; \Delta) \vdash^{-} A \wedge B} {\quad{(\Gamma; \Delta) \vdash^{-} B}} \quad \quad {\infer[\scriptstyle\wedge L^{c}]{(\Gamma';\Delta', A \wedge B) \vdash^{*} C} {{(\Gamma'; \Delta', A) \vdash^{*} C} \quad {(\Gamma'; \Delta', B) \vdash^{*} C}}}}}

and those can be transformed into derivations with cuts of cut-height $n+m$ or $n+k$, respectively:
\vspace{0.3cm}

  \quad
{\small
\infer[\scriptstyle Cut^{c}]{(\Gamma, \Gamma'; \Delta, \Delta') \vdash^{*} C}{{(\Gamma; \Delta) \vdash^{-} A} \quad \quad {(\Gamma';\Delta', A) \vdash^{*} C} }
\quad \quad
\infer[\scriptstyle Cut^{c}]{(\Gamma, \Gamma'; \Delta, \Delta') \vdash^{*} C}{{(\Gamma; \Delta) \vdash^{-} B} \quad \quad {(\Gamma';\Delta', B) \vdash^{*} C} }}

Here, both cut-height and weight of the cut formulas are reduced.

\item[-5.2-] $D = A \vee B$. The possible derivations for $Cut^{a}$ with a cut of cut-height $n+1+max(m, k)+1$ are
	\vspace{0.3cm}

  \quad
{\small
\infer[\scriptstyle Cut^{a}]{(\Gamma, \Gamma'; \Delta, \Delta') \vdash^{*} C}{\infer[\scriptstyle\vee R^{+}_{1}]{\quad \quad \quad (\Gamma; \Delta) \vdash^{+} A \vee B} {\quad{(\Gamma; \Delta) \vdash^{+} A}} \quad \quad {\infer[\scriptstyle\vee L^{a}]{(\Gamma', A \vee B;\Delta') \vdash^{*} C} {{(\Gamma', A; \Delta') \vdash^{*} C} \quad {(\Gamma', B; \Delta') \vdash^{*} C}}}}

or

\hspace{-0.1cm} 
\infer[\scriptstyle Cut^{a}]{(\Gamma, \Gamma'; \Delta, \Delta') \vdash^{*} C}{\infer[\scriptstyle\vee R^{+}_{2}]{\quad \quad \quad (\Gamma; \Delta) \vdash^{+} A \vee B} {\quad{(\Gamma; \Delta) \vdash^{+} B}} \quad \quad {\infer[\scriptstyle\vee L^{a}]{(\Gamma', A \vee B;\Delta') \vdash^{*} C} {{(\Gamma', A; \Delta') \vdash^{*} C} \quad {(\Gamma', B; \Delta') \vdash^{*} C}}}}}

and those can be transformed into derivations with cuts of cut-height $n+m$ and $n+k$, respectively:
\vspace{0.3cm}

  \quad
{\small
\infer[\scriptstyle Cut^{a}]{(\Gamma, \Gamma'; \Delta, \Delta') \vdash^{*} C}{{(\Gamma; \Delta) \vdash^{+} A} \quad \quad {(\Gamma', A;\Delta') \vdash^{*} C} }
\quad \quad
\infer[\scriptstyle Cut^{a}]{(\Gamma, \Gamma'; \Delta, \Delta') \vdash^{*} C}{{(\Gamma; \Delta) \vdash^{+} B} \quad \quad {(\Gamma', B;\Delta') \vdash^{*} C} }}

Again, both cut-height and weight of the cut formulas are reduced.

The derivation for $Cut^{c}$ with a cut of cut-height $max (n,m)+1+k+1$ is
	\vspace{0.3cm}

  \quad
{\small\infer[\scriptstyle Cut^{c}]{(\Gamma, \Gamma'; \Delta, \Delta') \vdash^{*} C}{\infer[\scriptstyle\vee R^{-}]{(\Gamma; \Delta) \vdash^{-} A \vee B} {{(\Gamma; \Delta) \vdash^{-} A} \quad {(\Gamma; \Delta) \vdash^{-} B}} \quad \quad 
{\infer[\scriptstyle\vee L^{c}]{(\Gamma'; \Delta', A \vee B) \vdash^{*} C} {{(\Gamma'; \Delta', A, B) \vdash^{*} C}}}}}   

and can be transformed into a derivation with two cuts of cut-height $n+k$ and $m+max(n,k)+1$:
\vspace{0.3cm}

  \quad
{\small
\infer[\scriptstyle C^{a/c}]{(\Gamma, \Gamma'; \Delta, \Delta') \vdash^{*} C} 
{\infer[\scriptstyle Cut^{c}]{(\Gamma, \Gamma, \Gamma'; \Delta, \Delta, \Delta') \vdash^{*} C}
{\quad \quad \quad {(\Gamma; \Delta) \vdash^{-} B} 
\quad \quad {\infer[\scriptstyle Cut^{c}]{(\Gamma, \Gamma'; \Delta, \Delta', B) \vdash^{*} C}{{(\Gamma; \Delta) \vdash^{-} A} \quad \quad {(\Gamma'; \Delta', A, B) \vdash^{*} C}}}}}
}

Note that again, in the case of the lower cut, although the cut-height might increase, the weight of the cut formula is reduced. 
For the upper cut both cut-height and weight of the cut formula is reduced.

\item[-5.3-] $D = A \rightarrow B$. The derivation for $Cut^{a}$ with a cut of cut-height $n+1+max(m, k)+1$ is
	\vspace{0.3cm}

  \quad
{\small
\infer[\scriptstyle Cut^{a}]{(\Gamma, \Gamma'; \Delta, \Delta') \vdash^{*} C}{\infer[\scriptstyle\rightarrow R^{+}]{\quad \quad \quad \quad(\Gamma; \Delta) \vdash^{+} A \rightarrow B} {{(\Gamma, A; \Delta) \vdash^{+} B}} \quad \quad {\infer[\scriptstyle\rightarrow L^{a}]{(\Gamma', A \rightarrow B;\Delta') \vdash^{*} C} {{(\Gamma', A \rightarrow B; \Delta') \vdash^{+} A} \quad {(\Gamma', B; \Delta') \vdash^{*} C}}}}
}

and this can be transformed into a derivation with three cuts of cut-height (from left to right and from top to bottom) $n+1+m$, $n+k$, and $max(n+1,m)+1+max(n,k)+1$ respectively:
\vspace{0.3cm}

  \quad
{\footnotesize
\infer[\scriptstyle C^{a/c}]{(\Gamma, \Gamma'; \Delta, \Delta') \vdash^{*} C}{\infer[\scriptstyle Cut^{a}]{(\Gamma, \Gamma, \Gamma', \Gamma'; \Delta, \Delta, \Delta', \Delta') \vdash^{*} C} {\infer[\scriptstyle Cut^{a}]{\quad \quad (\Gamma, \Gamma'; \Delta, \Delta') \vdash^{+} A} {\infer[\scriptstyle\rightarrow R^{+}]{(\Gamma; \Delta) \vdash^{+} A \rightarrow B} {(\Gamma, A; \Delta) \vdash^{+} B}  {(\Gamma', A \rightarrow B; \Delta') \vdash^{+} A}} \quad \quad {\infer[\scriptstyle Cut^{a}]{(\Gamma, A, \Gamma';\Delta, \Delta') \vdash^{*} C} {{(\Gamma, A; \Delta) \vdash^{+} B} \quad {(\Gamma', B; \Delta') \vdash^{*} C}}}}}
}

In the first case cut-height is reduced, in the second case cut-height and weight of the cut formula is reduced and in the third case weight of the cut formula is reduced.

The derivation for $Cut^{c}$ with a cut of cut-height $max(n,m)+1+k+1$ is
	\vspace{0.3cm}

  \quad
{\small
\infer[\scriptstyle Cut^{c}]{(\Gamma, \Gamma'; \Delta, \Delta') \vdash^{*} C}{\infer[\scriptstyle\rightarrow R^{-}]{(\Gamma; \Delta) \vdash^{-} A \rightarrow B} {{(\Gamma; \Delta) \vdash^{+} A}\quad {(\Gamma; \Delta) \vdash^{-} B}} \quad \quad {\infer[\scriptstyle\rightarrow L^{c}]{(\Gamma';\Delta', A \rightarrow B) \vdash^{*} C} {{(\Gamma', A; \Delta', B) \vdash^{*} C}}}}
}

This can be transformed into a derivation with two cuts of cut-height $n+k$ and $m+max(n,k)+1$:
\vspace{0.3cm}

  \quad
{\small
\infer[\scriptstyle C^{a/c}]{(\Gamma, \Gamma'; \Delta, \Delta') \vdash^{*} C} 
{\infer[\scriptstyle Cut^{c}]{(\Gamma, \Gamma, \Gamma'; \Delta, \Delta, \Delta') \vdash^{*} C}
{\quad \quad \quad {(\Gamma; \Delta) \vdash^{-} B} 
\quad \quad {\infer[\scriptstyle Cut^{a}]{(\Gamma, \Gamma'; \Delta, \Delta', B) \vdash^{*} C}{{(\Gamma; \Delta) \vdash^{+} A} \quad \quad {(\Gamma', A; \Delta', B) \vdash^{*} C}}}}}
}

In the first case cut-height and weight of the cut formula is reduced, while in the second case the weight of the cut formula is reduced.
Here we can observe a result specific for this calculus due to the mixture of derivability relations $\vdash^{+}$ and $\vdash^{-}$ in $\rightarrow R^{-}$ and the position of the active formulas in the assumptions \textit{and} in the counterassumptions in $\rightarrow L^{c}$:
Derivations containing instances of $Cut^{c}$ are not necessarily transformed into derivations with a lesser cut-height or a reduced weight of the cut formula of another instance of $Cut^{c}$ but it can also happen that $Cut^{c}$ is replaced by $Cut^{a}$.

\item[-5.4-] $D = A \Yleft B$. The derivation for $Cut^{a}$ with a cut of cut-height $max(n,m)+1+k+1$ is
	\vspace{0.3cm}

  \quad
{\small
\infer[\scriptstyle Cut^{a}]{(\Gamma, \Gamma'; \Delta, \Delta') \vdash^{*} C}{\infer[\scriptstyle\Yleft R^{+}]{(\Gamma; \Delta) \vdash^{+} A \Yleft B} {{(\Gamma; \Delta) \vdash^{+} A}\quad {(\Gamma; \Delta) \vdash^{-} B}} \quad \quad {\infer[\scriptstyle\Yleft L^{a}]{(\Gamma', A \Yleft B;\Delta') \vdash^{*} C} {{(\Gamma', A; \Delta', B) \vdash^{*} C}}}}
}

This can be transformed into a derivation with two cuts of cut-height $n+k$ and $m+max(n,k)+1$:
\vspace{0.3cm}

  \quad
{\small
\infer[\scriptstyle C^{a/c}]{(\Gamma, \Gamma'; \Delta, \Delta') \vdash^{*} C} 
{\infer[\scriptstyle Cut^{c}]{(\Gamma, \Gamma, \Gamma'; \Delta, \Delta, \Delta') \vdash^{*} C}
{\quad \quad \quad {(\Gamma; \Delta) \vdash^{-} B} 
\quad \quad {\infer[\scriptstyle Cut^{a}]{(\Gamma, \Gamma'; \Delta, \Delta', B) \vdash^{*} C}{{(\Gamma; \Delta) \vdash^{+} A} \quad \quad {(\Gamma', A; \Delta', B) \vdash^{*} C}}}}}
}

Again, due to the mixture of derivability relations $\vdash^{+}$ and $\vdash^{-}$ in $\Yleft R^{+}$ and the presence of the active formulas both in assumptions and counterassumptions in $\Yleft L^{a}$, in this case $Cut^{a}$ can be replaced by instances of $Cut^{c}$ with a reduced weight of the cut formula.
In the upper cut we have a reduction of both cut-height and weight of the cut formula.

The derivation for $Cut^{c}$ with a cut of cut-height $n+1+max(m, k)+1$ is
	\vspace{0.3cm}

  \quad
{\small
\infer[\scriptstyle Cut^{c}]{(\Gamma, \Gamma'; \Delta, \Delta') \vdash^{*} C}{\infer[\scriptstyle\Yleft R^{-}]{\quad \quad \quad \quad(\Gamma; \Delta) \vdash^{-} A \Yleft B} {{(\Gamma; \Delta, B) \vdash^{-} A}} \quad \quad {\infer[\scriptstyle\Yleft L^{c}]{(\Gamma';\Delta', A \Yleft B) \vdash^{*} C} {{(\Gamma'; \Delta', A \Yleft B) \vdash^{-} B} \quad {(\Gamma'; \Delta', A) \vdash^{*} C}}}}
}

and this can be transformed into a derivation with three cuts of cut-height (from left to right and from top to bottom) $n+1+m$, $n+k$, and $max(n+1,m)+1+max(n,k)+1$ respectively:
\vspace{0.3cm}

  \quad
{\footnotesize
\infer[\scriptstyle C^{a/c}]{(\Gamma, \Gamma'; \Delta, \Delta') \vdash^{*} C}{\infer[\scriptstyle Cut^{c}]{(\Gamma, \Gamma, \Gamma', \Gamma'; \Delta, \Delta, \Delta', \Delta') \vdash^{*} C} {\infer[\scriptstyle Cut^{c}]{\quad \quad (\Gamma, \Gamma'; \Delta, \Delta') \vdash^{-} B} {\infer[\scriptstyle\Yleft R^{-}]{(\Gamma; \Delta) \vdash^{-} A \Yleft B} {(\Gamma; \Delta, B) \vdash^{-} A}  {(\Gamma'; \Delta', A \Yleft B) \vdash^{-} B}} \quad \quad {\infer[\scriptstyle Cut^{c}]{(\Gamma, \Gamma';\Delta, B, \Delta') \vdash^{*} C} {{(\Gamma; \Delta, B) \vdash^{-} A} \quad {(\Gamma'; \Delta', A) \vdash^{*} C}}}}}
}

In the first case cut-height is reduced, in the second case cut-height and weight of the cut formula and in the third case weight of the cut formula.  
\end{itemize}
\qed

\section{Conclusion}
By applying the proof methods that \citet{Negri} use for their calculus \texttt{G3ip}, we were able to show that \texttt{SC2Int} is a cut-free sequent calculus for the bi-intuitionistic logic \texttt{2Int}. A proof can be given for the admissibility of the structural rules of weakening, contraction and cut in the system.

\end{document}